\documentclass[english,journal,draftclsnofoot,onecolumn]{IEEEtran}
\usepackage[T1]{fontenc}
\usepackage[latin9]{inputenc}
\usepackage{float}
\usepackage{amsmath}
\usepackage{amssymb}
\usepackage{stackrel}
\usepackage{graphicx}
\usepackage{setspace}
\makeatletter

\floatstyle{ruled}
\newfloat{algorithm}{tbp}{loa}
\providecommand{\algorithmname}{Algorithm}
\floatname{algorithm}{\protect\algorithmname}

\@ifundefined{date}{}{\date{}}
\@ifundefined{showcaptionsetup}{}{%
 \PassOptionsToPackage{caption=false}{subfig}}
\usepackage{subfig}
\makeatother

\usepackage{babel}
\begin{document}
\title{Designing unimodular sequence with good auto-correlation properties
via Block Majorization-Minimization method}
\author{Surya Prakash Sankuru, Prabhu Babu}
\maketitle
\begin{abstract}
Constant modulus sequence having lower side-lobe levels in its auto-correlation
function plays an important role in the applications like SONAR, RADAR
and digital communication systems. In this paper, we consider the
problem of minimizing the Integrated Sidelobe Level (ISL) metric,
to design a complex unimodular sequence of any length. The underlying
optimization problem is solved iteratively using the Block Majorization-Minimization
(MM) technique, which ensures that the resultant algorithm to be monotonic.
We also show a computationally efficient way to implement the algorithm
using Fast Fourier Transform (FFT) and Inverse Fast Fourier Transform
(IFFT) operations. Numerical experiments were conducted to compare
the proposed algorithm with the state-of-the art algorithms and was
found that the proposed algorithm performs better in terms of computational
complexity and speed of convergence.

Index Terms--Block Majorization-Minimization, Integrated Sidelobe
Level, unimodular sequence, aperiodic auto-correlation function, SONAR,
RADAR.
\end{abstract}

\section*{\centerline{I.INTRODUCTION AND PROBLEM FORMULATION}}

Transmit sequence with an impulse like aperiodic auto-correlation
function have many applications, e.g. high resolution SONAR imaging
\cite{1_DSP_SONAR}, \cite{7_probingwaveformrxsyn_robertHeStoica},
\cite{8_Probing_waveform}, RADAR imaging \cite{2_RadarHandbook},
\cite{3_Radarsignal_Levanon}, \cite{4_signal_App}, \cite{6_Phasecodedwaveform_Benedetto},
\cite{7_probingwaveformrxsyn_robertHeStoica} and CDMA communication
systems (to name a few) \cite{4_signal_App}, \cite{5_R_Turyn}, \cite{7_probingwaveformrxsyn_robertHeStoica},
\cite{9_waveform_textbook}. Hence, a sequence with lower side-lobe
levels in its auto-correlation function is usually desired. In addition
to minimizing side-lobe levels, we concentrate on the design of a
unimodular sequence due to the practical constraints such as usage
of full transmission power available in the system, avoidance of the
non-linear side effects and the limitations posed by sequence generation
hardware \cite{7_probingwaveformrxsyn_robertHeStoica}, \cite{9_waveform_textbook},
\cite{16_CAN}.

Let $\left\{ y_{i}\right\} _{i=1}^{N}$ be a complex unimodular sequence
of length $\text{\textquoteleft}N\text{\textquoteright}$ to be designed.
The aperiodic auto-correlation of a sequence $\left\{ y_{i}\right\} _{i=1}^{N}$
at any lag $\text{\textquoteleft}k\text{\textquoteright}$ is defined
as:

\begin{equation}
r(k)=\sum_{i=1}^{N-k}y_{i+k}y_{i}^{*}=r^{*}(-k),\hspace{1em}k=0,....,N-1.\label{eq:acor}
\end{equation}

There are two metrics, namely Integrated Side-lobe Level (ISL) and
Peak Side-lobe Level (PSL), which are commonly used to measure the
degree of correlation of a sequence. The ISL and PSL metrics of a
sequence are defined as:

\begin{equation}
\text{ISL}=\sum_{k=1}^{N-1}|r(k)|^{2}\label{eq:isl}
\end{equation}

\begin{equation}
\text{PSL}=\text{max}\left\{ |r(k)|\right\} _{k=1}^{N-1}\label{eq:psl}
\end{equation}

However, the ISL metric is usually peferred to design a sequence due
to its direct applicability to various applications. Hence, our problem
of interest would also be
\begin{equation}
\begin{aligned} & \underset{\boldsymbol{\boldsymbol{\boldsymbol{y}}}}{\text{\text{minimize}}} &  & \text{ISL}=\sum_{k=1}^{N-1}|r(k)|^{2}\\
 & \text{subject to} &  & |y_{i}|=1,\hspace{1em}i=1,...,N,
\end{aligned}
\label{eq:ISLprob}
\end{equation}
where $\boldsymbol{\boldsymbol{\boldsymbol{y}}}=[y_{1},y_{2},....,y_{N}]^{T}$.
The algorithms used to design unimodular sequences can be broadly
classified into two categories$-$analytical and computational. Some
of the sequences derived using analytical approach are Binary sequences
\cite{10_Binaryseq}, \cite{11_binaryseq}, \cite{12binary}, Frank
sequence \cite{13_Polyphasecode_frank}, Polyphase sequence \cite{14_polyphaseseq_borwein},
Golomb sequence \cite{15_polyphaseseq_Zhang}. But these sequences
exists only for limited length and has lesser degrees of freedom.
On the other hand, computational approaches are able to design a sequence
of arbitrary length but at the cost of high computational complexity.
Some of the computational approaches available in the literature are
CAN algorithm \cite{16_CAN}, MISL algorithm \cite{17_MISL}, ADMM
approach \cite{19_ADMM}, ISL-NEW algorithm \cite{20_fast_alg_waveform_LI}.

The following conventions for math symbols are adopted hereafter:
boldface uppercase letters denote matrices, boldface lowercase letters
denote column vectors and italics denote scalars. $\text{Tr}()$ denotes
the trace of a matrix. The superscripts $()^{T},()^{*},()^{H}$ denote
transpose, complex conjugate and conjugate transpose, respectively.
$\text{Re}(.)$ and $\text{Im}(.)$ denote real and imaginary parts,
respectively. $\text{arg}(.)$ denotes the phase of a complex number
and $y_{i}$ denote the $i^{th}$ element of vector $\boldsymbol{\boldsymbol{y}}$.
$\boldsymbol{\boldsymbol{I}}_{n}$ denotes the $n\times n$ identity
matrix and $\text{vec}(\boldsymbol{\boldsymbol{G}})$ is a column
vector consists of all the columns of a matrix-$\boldsymbol{\boldsymbol{G}}$
stacked. $\text{Diag}(\boldsymbol{\boldsymbol{y}})$ is a diagonal
matrix formed with $\boldsymbol{\boldsymbol{y}}$ as its diagonal.
$\left|.\right|^{2}$denotes the absolute squared value. $\mathbb{R}$
and $\mathbb{C}$ represent the real and complex fields. $\left\lfloor .\right\rfloor $
represents the nearest integer value.

CAN algorithm \cite{16_CAN} designs a sequence by minimizing an approximation
of the ISL function. The authors in \cite{16_CAN} rewrote the objective
function in (\ref{eq:ISLprob}) by expressing it in the frequency
domain as:

\begin{equation}
\sum_{k=1}^{N-1}|r(k)|^{2}=\frac{1}{4N}\sum_{f=1}^{2N}\Biggl[\Biggl|\sum_{i=1}^{N}y_{i}e^{-j\omega_{f}(i-1)}\Biggr|^{2}-N\Biggr]^{2}\label{eq:lag_freq}
\end{equation}

where $\omega_{f}=\frac{2\pi}{2N}(f-1),\,f=1,...,2N$ are the Fourier
grid frequencies.

Then the problem (\ref{eq:ISLprob}) can be rewritten as:
\begin{equation}
\begin{aligned} & \underset{\boldsymbol{\boldsymbol{\boldsymbol{y}}}}{\text{\text{minimize}}} &  & \frac{1}{4N}\sum_{f=1}^{2N}\Biggl[\Biggl|\sum_{i=1}^{N}y_{i}e^{-j\omega_{f}(i-1)}\Biggr|^{2}-N\Biggr]^{2}\\
 & \text{subject to} &  & |y_{i}|=1,\hspace{1em}i=1,...,N.
\end{aligned}
\label{eq:canprob}
\end{equation}

The cost function in (\ref{eq:canprob}) is a quartic function in
the variables $\left\{ y_{i}\right\} $ and it is hard to arrive at
a minimizer for (\ref{eq:canprob}). Thus, instead of solving (\ref{eq:canprob})
directly, the authors in \cite{16_CAN} solved an almost equivalent
problem, which has a quadratic cost function in $\left\{ y_{i}\right\} $
as shown below:

\begin{equation}
\begin{aligned} & \underset{\boldsymbol{\boldsymbol{\boldsymbol{y}}},\phi_{f}}{\text{\text{minimize}}} &  & \sum_{f=1}^{2N}\Biggl[\Biggl|\sum_{i=1}^{N}y_{i}e^{-j\omega_{f}(i-1)}-\sqrt{N}e^{j\phi_{f}}\Biggr|^{2}\Biggr]\\
 & \text{subject to} &  & |y_{i}|=1,\hspace{1em}i=1,...,N,
\end{aligned}
\label{eq:can2}
\end{equation}

where $\phi_{f}$$,\,f=1,2,...,2N$ are auxiliary variables.

The problem in (\ref{eq:can2}) can be rewritten as

\begin{equation}
\begin{aligned} & \underset{\boldsymbol{\boldsymbol{y}},\boldsymbol{\boldsymbol{x}}}{\text{\text{minimize}}} &  & \Biggl\Vert\boldsymbol{\boldsymbol{\hat{P}}}^{H}\boldsymbol{\boldsymbol{y}}-\sqrt{N}\boldsymbol{\boldsymbol{x}}\Biggr\Vert_{2}^{2}\\
 & \text{subject to} &  & |y_{i}|=1,\hspace{1em}i=1,...,N,
\end{aligned}
\label{eq:can3}
\end{equation}

where $\boldsymbol{\boldsymbol{\hat{P}}}=[\boldsymbol{\boldsymbol{p}}_{1},....,\boldsymbol{\boldsymbol{p}}_{2N}]$
be a $N\times2N$ matrix with $\boldsymbol{\boldsymbol{p}}_{f}\triangleq[1,e^{j\omega_{f}},...,e^{j\omega_{f}(N-1)}]^{T}$
and $\boldsymbol{\boldsymbol{x}}\triangleq[e^{j\phi_{1}},...,e^{j\phi_{2N}}]^{T}$.
CAN algorithm solves the problem (\ref{eq:can3}) by alternatively
minimizing between $\boldsymbol{\boldsymbol{y}}$ and $\boldsymbol{\boldsymbol{x}}$.
For a fixed $\boldsymbol{\boldsymbol{y}}$, minimization of (\ref{eq:can3})
with respect to $\phi_{f}$ is given by:

\begin{equation}
\phi_{f}=\text{arg}(u_{f}),\hspace{1em}f=1,..,2N,\label{eq:pfiupdate}
\end{equation}

where $\boldsymbol{\boldsymbol{u}}\triangleq\boldsymbol{\boldsymbol{\hat{P}}}^{H}\boldsymbol{\boldsymbol{y}}$
and for a fixed $\boldsymbol{\boldsymbol{x}}$, minimizer over $\boldsymbol{\boldsymbol{y}}$
would be:

\begin{equation}
y_{i}=e^{j\text{arg}(g_{i})},\hspace{1em}i=1,..,N,\label{eq:yupdate}
\end{equation}

where $\boldsymbol{\boldsymbol{g}}\triangleq\boldsymbol{\hat{\boldsymbol{P}}}\boldsymbol{\boldsymbol{x}}$.
The pseudocode of the CAN algorithm is summarized in the table Algorithm
1.

\begin{algorithm}[h]
\textbf{Require:} sequence length $\text{\textquoteleft}N\text{\textquoteright}$

1: set $t=0$, initialize $\boldsymbol{\boldsymbol{y}}^{0}$

2:\textbf{ repeat}

3: $\hphantom{n}$$\boldsymbol{\boldsymbol{u}}=\boldsymbol{\boldsymbol{\hat{P}}}^{H}\boldsymbol{\boldsymbol{y}}^{t}$

4: $\hphantom{n}$$x_{f}=e^{j\text{arg}(u_{f})},f=1,..,2N$

5: $\hphantom{n}$$\boldsymbol{\boldsymbol{g}}=\boldsymbol{\boldsymbol{\hat{P}x}}$

6: $\hphantom{n}$$y_{i}^{t+1}=e^{j\text{arg}(g_{i})},i=1,..,N$

7: $\hphantom{n}$$t$$\leftarrow$$t+1$

8:\textbf{ until} convergence

\caption{:The CAN algorithm proposed in \cite{16_CAN}}
\end{algorithm}

Since, CAN algorithm solves an approximation of the problem in (\ref{eq:canprob}),
the sequence obtained by solving the problem in (\ref{eq:can2}) will
not be a minimizer of the original problem in (\ref{eq:canprob}).
To fix this shortcoming, Song et.al. in \cite{17_MISL} proposed the
MISL algorithm by solving directly the problem in (\ref{eq:canprob}).
MISL solves the ISL minimization problem by MM method. Without going
into explanation of MISL algorithm, as it would require detailed explaination
of MM method, the pseudocode of the MISL algorithm summarized in the
table Algorithm 2.

\begin{algorithm}[h]
\textbf{Require:} sequence length $\text{\textquoteleft}N\text{\textquoteright}$

1: set $t=0$, initialize $\boldsymbol{\boldsymbol{y}}^{0}$

2:\textbf{ repeat}

3:$\hphantom{n}$ $\boldsymbol{\boldsymbol{u}}=\boldsymbol{\boldsymbol{\hat{P}}}^{H}\boldsymbol{\boldsymbol{y}}^{t}$

4:$\hphantom{n}$ $u_{\text{max}}=\text{max}{}_{f}\bigl\{|u_{f}|^{2}:f=1,..,2N\bigr\}$

5:$\hphantom{n}$ $\boldsymbol{\boldsymbol{z}}=-\boldsymbol{\boldsymbol{\hat{P}}}\Biggl(\text{Diag}\Bigl(\boldsymbol{|\boldsymbol{u}|^{2}}\Bigr)-u_{\text{max}}\boldsymbol{\boldsymbol{I}}-N^{2}\boldsymbol{\boldsymbol{I}}\Biggr)\boldsymbol{\boldsymbol{u}}$

6:$\hphantom{n}$ $y_{i}^{t+1}=e^{j\text{arg}(z_{i})},i=1,..,N$

7:$\hphantom{n}$ $t$$\leftarrow$$t+1$

8:\textbf{ until }convergence\caption{:The MISL algorithm proposed in \cite{17_MISL}}
\end{algorithm}

Even though MISL algorithm solves the original problem, it suffers
from slower speed of convergence. On the other hand, when compared
to CAN algorithm, it converges to the stationary point of problem
in (\ref{eq:canprob}). Both the algorithms are implemented via FFT
and IFFT operations and are computationally viable to be implemented
in standard pcs.

In \cite{18_MMcorr}, J. Song et.al solve the ISL metric problem to
design a sequence set and proposed an algorithm named as MM-Corr,
using the MM method. By taking number of sequences as one instead
of sequence set, observed that its performance is almost equal to
MISL algorithm. In \cite{19_ADMM}, J.Liang et.al proposed a new approach
to solve a problem in (\ref{eq:can3}) by using the ADMM method and
concludes that, such a technique had a poor performance when compared
to MISL algorithm interms of the PSL of an aperiodic auto-correlation
function. Y. Li et.al proposed the ISL-NEW algorithm \cite{20_fast_alg_waveform_LI}
by solving the problem in (\ref{eq:canprob}) by using MM method and
presented simulation results showing ISL-NEW algorithm as a faster
algorithm compared to MISL. J.song et.al had proposed an algorithm
based on the MM method named as MM-PSL \cite{21_MM_PSL}, by solving
$l_{p}$-norm of the auto-correlation function $\left(2<p<\infty\right)$
as an objective function, which is different from ISL metric.

The main motivation of this paper is to solve the original ISL minimization
problem in (\ref{eq:ISLprob}) with a better speed of convergence
(with lesser computational complexity) than the existing methods.
To achieve this, we have used Block MM technique. We also show a computationally
efficient way to implement our algorithm via FFT and IFFT operations.

The major contributions of the paper are as follows:
\begin{enumerate}
\item An algorithm based on the Block MM framework is proposed, to design
a sequence of any length $N$ by minimizing the ISL metric.
\item We also propose a computationally efficient way to implement our algorithm,
which we call as Fast Block MM (FBMM). This is particularly useful
for generating sequence of larger lengths.
\item We prove that the proposed algorithm converges to a stationary point
of the problem in (\ref{eq:ISLprob}).
\item Numerical experiments were conducted to prove that, our proposed algorithm
will perform better when compared to existing methods in terms of
speed of convergence.
\end{enumerate}
The rest of the paper is organised as follows. We first give an overview
of MM and Block-MM in section II. Next we propose our algorithm and
its faster version (FBMM) in section III and discuss its convergence
and computational complexity. Numerical experiments are discussed
in section IV and finally section V concludes the paper.

\section*{\centerline{II.Majorization-Minimization METHOD}}

\subsection*{A. MM Procedure:}

MM is an iterative procedure, which is used to solve an optimization
problem (non-convex or sometimes even a convex) more efficiently.
The MM procedure mainly consists of two steps with first step being
forming a surroagte function $g(\boldsymbol{\boldsymbol{y}}|\boldsymbol{\boldsymbol{y}}^{t})$
which majorizes (upper bounds) the original objective function $f(\boldsymbol{\boldsymbol{y}})$
at any feasible point $\boldsymbol{\boldsymbol{y}}=\boldsymbol{\boldsymbol{y}}^{t}$,
which is followed by minimizing the surrogate function to find the
next iterative estimate $\boldsymbol{\boldsymbol{y}}^{t+1}$. The
surrogate function $g(\boldsymbol{\boldsymbol{y}}|\boldsymbol{\boldsymbol{y}}^{t})$
has to satisfy the following properties:

\begin{equation}
g(\boldsymbol{\boldsymbol{y}}^{t}|\boldsymbol{\boldsymbol{y}}^{t})=f(\boldsymbol{\boldsymbol{y}}^{t}),\;\forall\boldsymbol{\boldsymbol{y}}\in\chi\label{eq:5}
\end{equation}

\begin{equation}
g(\boldsymbol{\boldsymbol{y}}|\boldsymbol{\boldsymbol{y}}^{t})\geq f(\boldsymbol{\boldsymbol{y}}),\;\forall\boldsymbol{\boldsymbol{y}}\in\chi\label{eq:6}
\end{equation}

where $\boldsymbol{\boldsymbol{y}}^{t}$ is the value taken by $\boldsymbol{\boldsymbol{y}}$
at $t^{th}$ iteration and $\chi$ is a set which consists all possible
values of $\boldsymbol{\boldsymbol{y}}$. Hence, the MM procedure
will generate the sequence of points $\left\{ \boldsymbol{\boldsymbol{y}}\right\} =\boldsymbol{\boldsymbol{y}}^{0},\boldsymbol{\boldsymbol{y}}^{1},\boldsymbol{\boldsymbol{y}}^{2},.....,\boldsymbol{\boldsymbol{y}}^{m}$
according to the following update rule:

\begin{equation}
\boldsymbol{\boldsymbol{y}}^{t+1}\triangleq\text{arg}\min_{\boldsymbol{\boldsymbol{y}\in\chi}}g(\boldsymbol{\boldsymbol{y}}|\boldsymbol{\boldsymbol{y}}^{t}).\label{eq:7}
\end{equation}

The objective value at every iteration will satisfy the following
descent property, i.e.
\begin{equation}
f(\boldsymbol{\boldsymbol{y}}^{t+1})\leq g(\boldsymbol{\boldsymbol{y}}^{t+1}|\boldsymbol{\boldsymbol{y}}^{t})\leq g(\boldsymbol{\boldsymbol{y}}^{t}|\boldsymbol{\boldsymbol{y}}^{t})=f(\boldsymbol{\boldsymbol{y}}^{t}).\label{eq:MM}
\end{equation}

Computational complexity and the convergence rate of MM based algorithms
mainly depends on the choice of the surrogate function $g(\boldsymbol{\boldsymbol{y}}|\boldsymbol{\boldsymbol{y}}^{t})$.
There are some guideline techniques to construct the surrogate functions
as discussed in \cite{22_MM_prabhubabu}, \cite{23_Tutorial_MM}.

\subsection*{B. Block MM:}

If one can split an optimization variable into $M$ blocks, then a
combination of Block Coordinate Descent \cite{24_Block_MM} and the
MM procedure can be applied i.e., the optimization variable is split
into blocks and then each block is treated as an independent variable
and updated using MM by keeping the other blocks fixed. Hence, the
$i^{th}$ block variable is updated by minimizing the surrogate function
$g_{i}(y_{i}|\boldsymbol{\boldsymbol{y}}^{t})$ which majorizes $f(y_{i})$
at a feasible point $\boldsymbol{\boldsymbol{y}}^{t}$ on the $i^{th}$
block. Such surrogate function has to satisfy the following properties:

\begin{equation}
\ensuremath{g_{i}(y_{i}^{t}|\boldsymbol{\boldsymbol{y}}^{t})=f(\boldsymbol{\boldsymbol{y}}^{t})},\label{eq:maj1}
\end{equation}

\begin{equation}
\ensuremath{g_{i}(y_{i}|\boldsymbol{\boldsymbol{y}}^{t})\geq f(y_{1}^{t},y_{2}^{t},..,y_{i},..,y_{N}^{t}),}\label{eq:maj2}
\end{equation}
where $\boldsymbol{\boldsymbol{y}}^{t}$ is the value taken by $\boldsymbol{\boldsymbol{y}}$
at the $t^{th}$ iteration.

The $i^{th}$ block at $(t+1)^{th}$ iteration is updated by solving
the following problem:

\begin{equation}
y_{i}^{t+1}\in\text{arg}\min_{y_{i}}g_{i}(y_{i}|\boldsymbol{\boldsymbol{y}}^{t}).\label{eq:maj3}
\end{equation}

In Block MM method, every block is updated in a sequential manner
and the surrogate function is choosen in a way, such that it is easy
to minimize and follow the shape of a objective function.

\section*{\centerline{III.ISL MINIMIZATION USING BLOCK MM TECHNIQUE}}

In this section, we present our algorithm and discuss its convergence
and computational complexity.

\subsection*{A. FBMM algorithm:}

Let us revisit the problem in (\ref{eq:ISLprob})

\[
\begin{aligned} & \underset{\boldsymbol{\boldsymbol{y}}}{\text{\text{minimize}}} &  & \sum_{k=1}^{N-1}|r(k)|^{2}\\
 & \text{subject to} &  & |y_{i}|=1,\hspace{1em}i=1,...,N.
\end{aligned}
\]

After substituting for $r(k)$, the above problem can be rewritten
as

\begin{equation}
\begin{aligned} & \underset{\boldsymbol{\boldsymbol{y}}}{\text{\text{minimize}}} &  & \Biggl|\sum_{i=1}^{N-1}y_{i+1}y_{i}^{*}\Biggr|^{2}+......+\Biggl|\sum_{i=1}^{2}y_{i+N-2}y_{i}^{*}\Biggr|^{2}+\Biggl|y_{N}y_{1}^{*}\Biggr|^{2}\\
 & \mbox{subject to} &  & |y_{i}|=1,\hspace{1em}i=1,...,N.
\end{aligned}
\label{eq:12}
\end{equation}

Now, to solve the problem in (\ref{eq:12}), we use the Block MM technique
by considering $y_{1},y_{2},..,y_{N}$ as an independent block variables.
For the sake of clarity, in the following we consider a generic optimization
problem in variable $y_{i}$, and optimization over any variable of
$\text{\textquotedblleft}\boldsymbol{y}\text{\textquotedblright}.$
would be very similar to the generic problem. Let the generic problem
be:

\begin{equation}
\begin{aligned} & \underset{y_{i}}{\text{\text{minimize}}} &  & f_{i}(y_{i})\\
 & \mbox{subject to} &  & |y_{i}|=1.
\end{aligned}
\label{eq:13}
\end{equation}

where $y_{i}$ indicates the $i^{th}$ block variable and its corresponding
objective function $f_{i}(y_{i})$ is defined as

\begin{equation}
f_{i}(y_{i})\triangleq a_{i}\Biggl[\sum_{k=1}^{l_{1}}\mid y_{i}m_{ki}^{*}+n_{ki}y_{i}^{*}+c_{ki}\mid^{2}\Biggr]+b_{i}\Biggl[\sum_{k=l_{2}}^{l_{3}}\mid n_{ki}y_{i}^{*}+c_{ki}\mid^{2}\Biggr]\label{eq:pre_gen_eq}
\end{equation}

where $a_{i},\,b_{i}$ are some fixed multiplicative constants, $l_{1},\,l_{2},\,l_{3}$
are the summation limits and $m_{ki},\,n_{ki},\,c_{ki}$ are the constants
associated with $k^{th}$ auto-correlation lag, which are given by

\begin{equation}
\begin{aligned} & m_{ki}\triangleq y_{i-k}\\
 & n_{ki}\triangleq y_{i+k}\\
 & c_{ki}\triangleq\sum_{q=k+1}^{N}(y_{q}y_{q-k}^{*}),\hphantom{m}q\neq i,q\neq k+i.
\end{aligned}
\label{eq:constants}
\end{equation}

The values that the variables $a_{i},\,b_{i},\,l_{1},\,l_{2},\,l_{3}$
take will depend on the variable index $(y_{i})$. They can be given
as follows:
\begin{flushleft}
\begin{equation}
\begin{aligned} & a_{i}\triangleq\begin{cases}
0 & i=1,N\\
1 & else
\end{cases}\hphantom{nn},\forall N.\\
 & b_{i}\triangleq\begin{cases}
1 & \forall i\end{cases},\forall N\in even.\\
 & b_{i}\triangleq\begin{cases}
0 & i=\left\lfloor N/2\right\rfloor +1\\
1 & else
\end{cases}\hphantom{nn},\forall N\in odd.
\end{aligned}
\label{eq:const}
\end{equation}
\par\end{flushleft}

\begin{equation}
\begin{aligned} & l_{1}\triangleq\begin{cases}
i-1 & i=2,..,\left\lfloor N/2\right\rfloor ,a_{i}\neq0\hphantom{nn},\forall N.\\
i-1 & b_{i}=0,a_{i}\neq0\\
N-i & i=\left\lfloor N/2\right\rfloor +1,a_{i}\neq0\hphantom{nn},\forall N\in even.\\
N-i & i=\left\lfloor N/2\right\rfloor +2,..,N-1,a_{i}\neq0\hphantom{nn},\forall N.
\end{cases}\\
 & l_{2}\triangleq\begin{cases}
i & a_{i}=0\\
l_{1}+1 & b_{i}\neq0\hphantom{nn},\forall N.
\end{cases}\\
 & l_{3}\triangleq\begin{cases}
N-1 & a_{i}=0\\
N-i & i=2,..,\left\lfloor N/2\right\rfloor \hphantom{nn},\forall N.\\
i-1 & b_{i}\neq0
\end{cases}
\end{aligned}
\label{eq:limits}
\end{equation}

So, from (\ref{eq:pre_gen_eq}), we have

\[
f_{i}(y_{i})=a_{i}\Biggl[\sum_{k=1}^{l_{1}}\mid y_{i}m_{ki}^{*}+n_{ki}y_{i}^{*}+c_{ki}\mid^{2}\Biggr]+b_{i}\Biggl[\sum_{k=l_{2}}^{l_{3}}\mid n_{ki}y_{i}^{*}+c_{ki}\mid^{2}\Biggr].
\]

which can be rewritten as

\begin{equation}
f_{i}(y_{i})=a_{i}\Biggl[\sum_{k=1}^{l_{1}}\biggl|y_{i}m_{ki}^{*}+n_{ki}y_{i}^{*}+c_{ki}\biggr|^{2}\Biggr]+b_{i}\Biggl[\sum_{k=l_{2}}^{l_{3}}\biggl|n_{ki}+c_{ki}y_{i}\biggr|^{2}\Biggr]\label{eq:21}
\end{equation}

Further simplification yields:

\begin{equation}
f_{i}(y_{i})=a_{i}\Biggl[\sum_{k=1}^{l_{1}}\biggl|y_{i}m_{ki}^{*}+n_{ki}y_{i}^{*}+c_{ki}\biggr|^{2}\Biggr]+b_{i}\Biggl[\sum_{k=l_{2}}^{l_{3}}w_{ki}\biggl|y_{i}+d_{ki}\biggr|^{2}\Biggr]\label{eq:22}
\end{equation}

where
\[
d_{ki}\triangleq\frac{n_{ki}}{c_{ki}},\,w_{ki}\triangleq|c_{ki}|^{2}.
\]

Expanding the square term in (\ref{eq:22}) and by ignoring the constant
terms, (\ref{eq:22}) can be rewritten as

\begin{equation}
\begin{array}{c}
f_{i}(y_{i})=\stackrel[k=1]{l_{1}}{\sum}a_{i}\biggl[(n_{ki}^{*}m_{ki}^{*})(y_{i}^{2})+(c_{ki}^{*}m_{ki}^{*}+n_{ki}^{*}c_{ki})(y_{i})+(n_{ki}m_{ki})(y_{i}^{2})^{*}+(m_{ki}c_{ki}+c_{ki}^{*}n_{ki})(y_{i})^{*}\biggr]\\
+\stackrel[k=l_{2}]{l_{3}}{\sum}b_{i}w_{ki}\Bigl[y_{i}d_{ki}^{*}+d_{ki}y_{i}^{*}\Bigr]
\end{array}\label{eq:24-1}
\end{equation}

\begin{equation}
f_{i}(y_{i})=\sum_{k=1}^{l_{1}}a_{i}\biggl[2\text{Re}\Bigl((n_{ki}^{*}m_{ki}^{*})(y_{i}^{2})\Bigr)+2\text{Re}\Bigl((c_{ki}^{*}m_{ki}^{*}+n_{ki}^{*}c_{ki})(y_{i})\Bigr)\biggr]+\sum_{k=l_{2}}^{l_{3}}\biggl[b_{i}w_{ki}*2\text{Re}\Bigl(y_{i}d_{ki}^{*}\Bigr)\biggr]\label{eq:26}
\end{equation}

\begin{flushleft}
Now if we define:
\begin{equation}
\begin{aligned} & n_{ki}^{*}m_{ki}^{*}\triangleq\hat{a}_{1ki}+j\hat{a}_{2ki}\\
 & (c_{ki}^{*}m_{ki}^{*}+n_{ki}^{*}c_{ki})\triangleq\hat{b}_{1ki}+j\hat{b}_{2ki}\\
 & d_{ki}^{*}\triangleq\hat{c}_{1ki}+j\hat{c}_{2ki}\\
 & y_{i}\triangleq u_{1}+ju_{2}
\end{aligned}
\label{eq:convert}
\end{equation}
\par\end{flushleft}

where $\hat{a}_{1ki},\,\hat{a}_{2ki},\,\hat{b}_{1ki},\,\hat{b}_{2ki},\,\hat{c}_{1ki},\,\hat{c}_{2ki},\,u_{1},\,u_{2}$
are real valued quantities.
\begin{flushleft}
Then $f_{i}(y_{i})$ in (\ref{eq:26}) can be further simplified as:
\par\end{flushleft}

\begin{equation}
\begin{array}{c}
f_{i}(u_{1},u_{2})=\Biggl[2a_{i}\stackrel[k=1]{l_{1}}{\sum}\hat{a}_{1ki}\Biggr](u_{1})^{2}-\Biggl[4a_{i}\stackrel[k=1]{l_{1}}{\sum}\hat{a}_{2ki}\Biggr]u_{1}u_{2}+\Biggl[2a_{i}\stackrel[k=1]{l_{1}}{\sum}\hat{b}_{1ki}+2b_{i}\stackrel[k=l_{2}]{l_{3}}{\sum}w_{ki}\hat{c}_{1ki}\Biggr]u_{1}\\
-\Biggl[\stackrel[k=1]{l_{1}}{\sum}2a_{i}\hat{b}_{2ki}+\stackrel[k=l_{2}]{l_{3}}{\sum}2b_{i}w_{ki}\hat{c}_{2ki}\Biggr]u_{2}-\Biggl[2a_{i}\stackrel[k=1]{l_{1}}{\sum}\hat{a}_{1ki}\Biggr](u_{2})^{2}
\end{array}\label{eq:solso-1}
\end{equation}

\begin{flushleft}
Again introducing,
\begin{equation}
\begin{aligned} & a\triangleq2a_{i}\sum_{k=1}^{l_{1}}\hat{a}_{1ki}\\
 & b\triangleq4a_{i}\sum_{k=1}^{l_{1}}(\hat{a}_{2ki})\\
 & c\triangleq2a_{i}\sum_{k=1}^{l_{1}}\hat{b}_{1ki}+2b_{i}\sum_{k=l_{2}}^{l_{3}}w_{ki}\hat{c}_{1ki}\\
 & d\triangleq2a_{i}\sum_{k=1}^{l_{1}}\hat{b}_{2ki}+2b_{i}\sum_{k=l_{2}}^{l_{3}}w_{ki}\hat{c}_{2ki}
\end{aligned}
\label{eq:genera}
\end{equation}
\par\end{flushleft}

\begin{flushleft}
Then $f_{i}(u_{1},u_{2})$ in (\ref{eq:solso-1}) is simplified as:
\par\end{flushleft}

\begin{equation}
f_{i}(u_{1},u_{2})=au_{1}^{2}-bu_{1}u_{2}+cu_{1}-du_{2}-au_{2}^{2}\label{eq:31}
\end{equation}
Thus the problem in (\ref{eq:13}) has become the following problem
with real valued variables.

\begin{equation}
\begin{aligned} & \underset{u_{1},u_{2}}{\text{\text{minimize}}} &  & f_{i}(u_{1},u_{2})\\
 & \text{subject to} &  & u_{1}^{2}+u_{2}^{2}=1.
\end{aligned}
\label{eq:32}
\end{equation}

\begin{flushleft}
Now, the problem in (\ref{eq:32}) can be written in (matrix-vector)
form as:
\par\end{flushleft}

\begin{equation}
\begin{aligned} & \underset{\boldsymbol{\boldsymbol{v}}}{\text{\text{minimize}}} &  & \boldsymbol{\boldsymbol{v}}^{T}\boldsymbol{\boldsymbol{A}}\boldsymbol{\boldsymbol{v}}+\boldsymbol{\boldsymbol{e}}^{T}\boldsymbol{\boldsymbol{v}}\\
 & \text{subject to} &  & \boldsymbol{\boldsymbol{\boldsymbol{v}}}^{T}\boldsymbol{\boldsymbol{\boldsymbol{v}}}=1,
\end{aligned}
\label{eq:33}
\end{equation}

with
\begin{equation}
\begin{aligned} & \boldsymbol{\boldsymbol{A}}\triangleq\begin{bmatrix}a & \frac{-b}{2}\\
\frac{-b}{2} & -a
\end{bmatrix}\\
 & \boldsymbol{\boldsymbol{e}}\triangleq\begin{bmatrix}c\\
-d
\end{bmatrix}\\
 & \boldsymbol{\boldsymbol{v}}\triangleq\begin{bmatrix}u_{1}\\
u_{2}
\end{bmatrix}
\end{aligned}
\label{eq:matrix}
\end{equation}

The problem in (\ref{eq:33}) has an objective function which is a
non-convex quadratic function in the variable $\boldsymbol{\boldsymbol{\boldsymbol{v}}}$
because of $(-a)$ in the diagonal of $\boldsymbol{\boldsymbol{A}}$
and also the constraint is a quadratic equality constraint, so the
problem in (\ref{eq:33}) is a non convex problem and hard to solve.
So, we decided to employ MM technique to solve the problem in (\ref{eq:33}).
Let us introduce the following lemma, which would be useful to develop
our algorithm.

\textbf{Lemma-1}: Let \textbf{$\boldsymbol{\boldsymbol{Q}}$ }be an
$n \times n$ Hermitian matrix and \textbf{$\boldsymbol{\boldsymbol{R}}$
}be another $n \times n$ Hermitian matrix such that $\boldsymbol{\boldsymbol{R}}\geq\boldsymbol{\boldsymbol{Q}}$.
Then for any point $\boldsymbol{\boldsymbol{\boldsymbol{y}}}^{0}\in\boldsymbol{C}^{n}$,
the quadratic function $\boldsymbol{\boldsymbol{\boldsymbol{y}}}^{H}\boldsymbol{\boldsymbol{\boldsymbol{Q}}}\boldsymbol{\boldsymbol{\boldsymbol{y}}}$
is majorized by $\boldsymbol{\boldsymbol{\boldsymbol{y}}}^{H}\boldsymbol{\boldsymbol{\boldsymbol{R}}}\boldsymbol{\boldsymbol{\boldsymbol{y}}}+2Re(\boldsymbol{\boldsymbol{\boldsymbol{y}}}^{H}(\boldsymbol{\boldsymbol{\boldsymbol{Q}}-\boldsymbol{\boldsymbol{R}}})\boldsymbol{\boldsymbol{\boldsymbol{y}}}^{0})+(\boldsymbol{\boldsymbol{\boldsymbol{y}}}^{0})^{H}(\boldsymbol{\boldsymbol{\boldsymbol{R}}-\boldsymbol{\boldsymbol{Q}}})\boldsymbol{\boldsymbol{\boldsymbol{y}}}^{0}$
at $\boldsymbol{\boldsymbol{\boldsymbol{y}}}^{0}$.

Proof: Although the proof can be find in \cite{17_MISL}, we replicate
it here for the sake of clarity.

As $\boldsymbol{\boldsymbol{\boldsymbol{R}}}\geq\boldsymbol{\boldsymbol{\boldsymbol{Q}}}$,
we have

$\begin{aligned}\boldsymbol{\boldsymbol{\boldsymbol{y}}}^{H}\boldsymbol{\boldsymbol{Q}}\boldsymbol{\boldsymbol{\boldsymbol{y}}} & =(\boldsymbol{\boldsymbol{\boldsymbol{y}}}^{0})^{H}\boldsymbol{\boldsymbol{\boldsymbol{Q}}}\boldsymbol{\boldsymbol{\boldsymbol{y}}}^{0}+2Re((\boldsymbol{\boldsymbol{\boldsymbol{y}}}-\boldsymbol{\boldsymbol{y}}^{0})^{H}\boldsymbol{\boldsymbol{\boldsymbol{Q}}}\boldsymbol{\boldsymbol{\boldsymbol{y}}}^{0})+(\boldsymbol{\boldsymbol{\boldsymbol{y}}}-\boldsymbol{\boldsymbol{\boldsymbol{y}}}^{0})^{H}\boldsymbol{\boldsymbol{\boldsymbol{Q}}}(\boldsymbol{\boldsymbol{\boldsymbol{y}}}-\boldsymbol{\boldsymbol{\boldsymbol{y}}}^{0})\\
 & \leq(\boldsymbol{\boldsymbol{\boldsymbol{y}}}^{0})^{H}\boldsymbol{\boldsymbol{\boldsymbol{Q}}}\boldsymbol{\boldsymbol{\boldsymbol{y}}}^{0}+2Re((\boldsymbol{\boldsymbol{\boldsymbol{y}}}-\boldsymbol{\boldsymbol{\boldsymbol{y}}}^{0})^{H}\boldsymbol{\boldsymbol{\boldsymbol{Q}}}\boldsymbol{\boldsymbol{y}}^{0})+(\boldsymbol{\boldsymbol{\boldsymbol{y}}}-\boldsymbol{\boldsymbol{\boldsymbol{y}}}^{0})^{H}\boldsymbol{\boldsymbol{\boldsymbol{R}}}(\boldsymbol{\boldsymbol{\boldsymbol{y}}}-\boldsymbol{\boldsymbol{\boldsymbol{y}}}^{0})\\
 & =\boldsymbol{\boldsymbol{\boldsymbol{y}}}{}^{H}\boldsymbol{\boldsymbol{\boldsymbol{R}}}\boldsymbol{\boldsymbol{\boldsymbol{y}}}+2Re(\boldsymbol{\boldsymbol{\boldsymbol{y}}}^{H}(\boldsymbol{\boldsymbol{\boldsymbol{Q}}-\boldsymbol{\boldsymbol{R}}})\boldsymbol{\boldsymbol{\boldsymbol{y}}}^{0})+(\boldsymbol{\boldsymbol{\boldsymbol{y}}}^{0})^{H}(\boldsymbol{\boldsymbol{\boldsymbol{R}}-\boldsymbol{\boldsymbol{Q}}})\boldsymbol{\boldsymbol{\boldsymbol{y}}}^{0}
\end{aligned}
$

for any $\boldsymbol{\boldsymbol{\boldsymbol{y}}}\in\boldsymbol{C}^{n}$.$\hphantom{nnnnnnnnnnnnnnnnnnnnnnnnnnnnnnnnnnnnnnnnnnnnnnnnnnnnnnnnnnnnnnn}$$\blacksquare$

Now, by using Lemma-1, we will majorize only the quadratic term in
the objective function of problem in (\ref{eq:33}) at any feasible
point $\boldsymbol{\boldsymbol{v}}=\boldsymbol{\boldsymbol{v}}^{t}$
and get

\begin{equation}
g_{i}(\boldsymbol{\boldsymbol{\boldsymbol{v}}}|\boldsymbol{\boldsymbol{\boldsymbol{v}}}^{t})=\boldsymbol{\boldsymbol{v}}^{T}\boldsymbol{\boldsymbol{\boldsymbol{A}}_{1}}\boldsymbol{\boldsymbol{v}}+2\text{Re}[\boldsymbol{\boldsymbol{v}}^{T}(\boldsymbol{\boldsymbol{\boldsymbol{A}}}-\boldsymbol{\boldsymbol{\boldsymbol{A}}_{1}})\boldsymbol{\boldsymbol{v}}^{t}]+(\boldsymbol{\boldsymbol{v}}^{t})^{T}(\boldsymbol{\boldsymbol{\boldsymbol{A}}_{1}}-\boldsymbol{\boldsymbol{\boldsymbol{A}}})\boldsymbol{\boldsymbol{v}}^{t},\label{eq:35}
\end{equation}

where $\boldsymbol{\boldsymbol{\boldsymbol{A}}_{1}}=\lambda_{\text{max}}(\boldsymbol{\boldsymbol{\boldsymbol{A}}}).\boldsymbol{\boldsymbol{I}}_{n}$
. Since $\lambda_{\text{max}}(\boldsymbol{\boldsymbol{\boldsymbol{A}}})$
is a constant value and $\boldsymbol{\boldsymbol{v}}^{T}\boldsymbol{\boldsymbol{v}}=1$,
so the first and the last terms in the above surrogate function are
constants. Hence, after ignoring the constant terms from (\ref{eq:35})
we get,

\begin{equation}
g_{i}(\boldsymbol{\boldsymbol{\boldsymbol{v}}}|\boldsymbol{\boldsymbol{\boldsymbol{v}}}^{t})=2\text{Re}[\boldsymbol{\boldsymbol{v}}^{T}(\boldsymbol{\boldsymbol{\boldsymbol{A}}}-\boldsymbol{\boldsymbol{\boldsymbol{A}}_{1}})\boldsymbol{\boldsymbol{v}}^{t}]\label{eq:surr_MODIFY}
\end{equation}
Now, the problem (\ref{eq:33}) is equal to

\begin{equation}
\begin{aligned} & \underset{\boldsymbol{\boldsymbol{\boldsymbol{v}}}}{\text{\text{minimize}}} &  & 2\text{Re}[\boldsymbol{\boldsymbol{v}}^{T}(\boldsymbol{\boldsymbol{\boldsymbol{A}}}-\boldsymbol{\boldsymbol{\boldsymbol{A}}_{1}})\boldsymbol{\boldsymbol{v}}^{t}]+\boldsymbol{\boldsymbol{e}}^{T}\boldsymbol{\boldsymbol{v}}\\
 & \text{subject to} &  & \boldsymbol{\boldsymbol{\boldsymbol{v}}}^{T}\boldsymbol{\boldsymbol{\boldsymbol{v}}}=1
\end{aligned}
\label{eq:36}
\end{equation}

which can be further rewritten as
\begin{equation}
\begin{aligned} & \underset{\boldsymbol{\boldsymbol{v}}}{\text{\text{minimize}}} &  & g_{i}(\boldsymbol{\boldsymbol{\boldsymbol{v}}}|\boldsymbol{\boldsymbol{\boldsymbol{v}}}^{t})=\parallel\boldsymbol{\boldsymbol{v}}-\boldsymbol{\boldsymbol{\boldsymbol{z}}}\parallel_{2}^{2}\\
 & \text{subject to} &  & \boldsymbol{\boldsymbol{\boldsymbol{v}}}^{T}\boldsymbol{\boldsymbol{\boldsymbol{v}}}=1
\end{aligned}
\label{eq:37}
\end{equation}

where $\boldsymbol{\boldsymbol{z}}=-[(\boldsymbol{\boldsymbol{\boldsymbol{A}}}-\boldsymbol{\boldsymbol{\boldsymbol{A}}_{1}})\boldsymbol{\boldsymbol{v}}^{t}+(\boldsymbol{\boldsymbol{e}}/2)]$.

Now, the problem in (\ref{eq:37}) has a closed form solution of:
\begin{equation}
\boldsymbol{\boldsymbol{\boldsymbol{v}}}=\frac{\boldsymbol{\boldsymbol{z}}}{||\boldsymbol{\boldsymbol{z}}||_{2}}.\label{eq:38}
\end{equation}

Then the update $y_{i}^{t+1}$ can be calculated by:

\begin{equation}
y_{i}^{t+1}=v_{1}+jv_{2}\label{eq:39}
\end{equation}

The constants $(c_{1i},c_{2i},...,c_{(N-1)i})$ in (\ref{eq:constants})
which are calculated at every iteration, which form the bulk of the
computations, can be computed via FFT and IFFT operations as follows:
For example, the constant $(c_{1i})$ can be intreperated as the auto-correlation
of a sequence $(\text{with }y_{i}=0)$ which inturn can be calculated
by an FFT and IFFT operation. So, to calculate all the constants of
$N$ variables, we would require $N$ number of FFT and $N$ number
of IFFT operations. To avoid implementing FFT and IFFT operations
$N$ number of times, we propose an computationally efficient way
to calculate the constants. To achieve this, we would exploit the
cyclic pattern in the expression of the constants. First we define
$\boldsymbol{\boldsymbol{s}}$ which includes original variable $\boldsymbol{\boldsymbol{y}}$
along with some pre-defined zero padding structure as shown below:

\begin{equation}
\boldsymbol{\boldsymbol{s}}=[\boldsymbol{0}_{1\times N-2},\boldsymbol{\boldsymbol{y}}^{T},\boldsymbol{0}_{1\times N}]^{T}\label{eq:newy}
\end{equation}

Let us define the variables $\boldsymbol{\boldsymbol{b}}_{i}$ and
$\boldsymbol{\boldsymbol{D}}_{i}$ as:

\begin{equation}
\boldsymbol{\boldsymbol{b}}_{i}=[-y_{i}^{*},y_{i-1}^{*},-y_{i},y_{i-1}]^{T}\label{eq:vecb}
\end{equation}

\begin{equation}
\boldsymbol{\boldsymbol{D}}_{i}=\begin{bmatrix}\boldsymbol{\boldsymbol{\boldsymbol{s}}}_{(N+i-1)} & . & . & \boldsymbol{\boldsymbol{\boldsymbol{s}}}_{(2N+i-4)} & 0\\
0 & \boldsymbol{\boldsymbol{\boldsymbol{s}}}_{(N+i-1)} & . & . & \boldsymbol{\boldsymbol{s}}_{(2N+i-4)}\\
0 & \boldsymbol{\boldsymbol{\boldsymbol{s}}}_{(N+i-4)}^{*} & . & . & \boldsymbol{\boldsymbol{\boldsymbol{s}}}_{(i-1)}^{*}\\
\boldsymbol{\boldsymbol{\boldsymbol{s}}}_{(N+i-4)}^{*} & . & . & \boldsymbol{\boldsymbol{\boldsymbol{s}}}_{(i-1)}^{*} & 0
\end{bmatrix}\label{eq:matD}
\end{equation}

So, to calculate the $i^{th}$ variable constants ($c_{1i},c_{2i},..,c_{(N-1)i})$,
we will use the constants associated with the $(i-1)^{th}$ variable
($c_{1(i-1)},c_{2(i-1)},..,c_{(N-1)(i-1)})$ as follows:

\begin{equation}
\begin{aligned}\begin{bmatrix}c_{1i}, & . & . & ,c_{(N-1)i}\end{bmatrix}= & \begin{bmatrix}c_{1(i-1)},. & . & . & ,c_{(N-1)(i-1)}\end{bmatrix}+\boldsymbol{\boldsymbol{b}}_{i}^{T}\boldsymbol{\boldsymbol{D}}_{i}\end{aligned}
\hphantom{nnn}\forall\,i=2,..,N\label{eq:constpatt}
\end{equation}
 Therefore, all the $(N-1)$ number of constants associated with each
of the $N$ variables are implemented using only one FFT and IFFT
operation. The steps of our algorithm which is named as FBMM is shown
in the table Algorithm 3.

\begin{algorithm}[h]
\textbf{Require:} sequence length $\text{\textquoteleft}N\text{\textquoteright}$

1: set $t=0$, initialize $\boldsymbol{y}^{0}$

2:\textbf{ repeat}

3:$\hphantom{nn}$\textbf{ }set $i=1$

4:$\hphantom{nnnnn}$\textbf{ repeat}

5:$\hphantom{nnnnnnnn}$calculate $\left\{ c_{ki}\right\} _{k=1}^{N-1}$
using (\ref{eq:constpatt})

6:$\hphantom{nnnnnnnn}$calculate $d_{ki}=\frac{n_{ki}}{c_{ki}},\,w_{ki}=|c_{ki}|^{2}\;,k=1,...,N-1.$

7:$\hphantom{nnnnnnnn}$$\boldsymbol{\boldsymbol{\boldsymbol{A}}}_{1}$=$\lambda_{\text{max}}(\boldsymbol{\boldsymbol{A}})$.$\boldsymbol{I}_{2}$

8:$\hphantom{nnnnnnnn}$ $\boldsymbol{\boldsymbol{z}}=-[(\boldsymbol{\boldsymbol{\boldsymbol{A}}}-\boldsymbol{\boldsymbol{\boldsymbol{A}}_{1}})\boldsymbol{\boldsymbol{v}}^{t}+(\boldsymbol{\boldsymbol{e}}/2)]$

9:$\hphantom{nnnnnnnn}$$\boldsymbol{\boldsymbol{\boldsymbol{v}}}=\frac{\boldsymbol{\boldsymbol{z}}}{||\boldsymbol{\boldsymbol{\boldsymbol{z}}}||_{2}}$

10:$\hphantom{nnnnnnn}$ $y_{i}^{t+1}=v_{1}+jv_{2}$

11:$\hphantom{nnnnnnn}$$i\longleftarrow i+1$

12:$\hphantom{nnnnn}$\textbf{until }length of a sequence

13:$\hphantom{nn}$$t\longleftarrow t+1$

14:\textbf{ until} convergence

\caption{:FBMM\textbf{ }algorithm}
\end{algorithm}

\subsection*{B. Proof of convergence:}

The proposed algorithm is based on a Block MM technique. As Block
MM is a combination of coordinate descent and the MM procedure, it
is ensured that the cost function evaluated at every limit point is
monotonic. Also, since the cost function in (\ref{eq:ISLprob}) is
bounded below by zero, the sequence of objective values is guaranteed
to converge to a finite value. In \cite{25_BMM_convergence}, Theorem
2.a Razaviyayn et.al stated that a limit point generated at each iteration
by a Block MM algorithm is a coordinate wise minimum point with respect
to original cost function, iff the upper bound $g_{i}(.)$ is a quasi-convex
function. We now have to prove that $g_{i}(\boldsymbol{\boldsymbol{\boldsymbol{v}}}|\boldsymbol{\boldsymbol{\boldsymbol{v}}}^{t})$
in (\ref{eq:35}) is indeed a quasi convex function.

So, from (\ref{eq:35}), we have,

\[
g_{i}(\boldsymbol{\boldsymbol{\boldsymbol{v}}}|\boldsymbol{\boldsymbol{\boldsymbol{v}}}^{t})=\boldsymbol{\boldsymbol{v}}^{T}\boldsymbol{\boldsymbol{\boldsymbol{A}}_{1}}\boldsymbol{\boldsymbol{v}}+2\text{Re}[\boldsymbol{\boldsymbol{v}}^{T}(\boldsymbol{\boldsymbol{\boldsymbol{A}}}-\boldsymbol{\boldsymbol{\boldsymbol{A}}_{1}})\boldsymbol{\boldsymbol{v}}^{t}]+(\boldsymbol{\boldsymbol{v}}^{t})^{T}(\boldsymbol{\boldsymbol{\boldsymbol{A}}_{1}}-\boldsymbol{\boldsymbol{\boldsymbol{A}}})\boldsymbol{\boldsymbol{v}}^{t}
\]

which is a quadratic function in $\boldsymbol{\boldsymbol{\boldsymbol{v}}}$.
The Hessian of $g_{i}(\boldsymbol{\boldsymbol{\boldsymbol{v}}}|\boldsymbol{\boldsymbol{\boldsymbol{v}}}^{t})$
is $2\boldsymbol{\boldsymbol{A}}_{1}$, where $\boldsymbol{\boldsymbol{A}}_{1}=\lambda_{\text{max}}(\boldsymbol{\boldsymbol{A}}).\boldsymbol{\boldsymbol{I}}_{n}$.
Since $\lambda_{\text{max}}(\boldsymbol{\boldsymbol{A}})$ is a positive
value, $\boldsymbol{\boldsymbol{A}}_{1}$ is a diagonal matrix with
positive entries. Hence $g_{i}(\boldsymbol{\boldsymbol{\boldsymbol{v}}}|\boldsymbol{\boldsymbol{\boldsymbol{v}}}^{t})$
is a convex function. Since every convex function is also a quasi-convex
function, $g_{i}(\boldsymbol{\boldsymbol{\boldsymbol{v}}}|\boldsymbol{\boldsymbol{\boldsymbol{v}}}^{t})$
is also a quasi-convex function. Therefore, according to Theorem 2.a
of \cite{25_BMM_convergence} the sequence of points generated by
FBMM will converge to the stationary point of problem in (\ref{eq:ISLprob}).

\subsection*{C. Computational complexity:}

The per iteration computational complexity of the proposed algorithm
is dominated in the calculation of constants $c_{ki}\;,k=1,..,N-1$
,$i=1,..,N$. These constants can be calculated using one FFT and
IFFT operation and the approach as mentioned in the end of subsection
(A), where we exploit some cyclic pattern and calculate the constants,
then the computational complexity per iteration would be $\mathcal{O}(N^{2})+\mathcal{O}(N\,logN)$.

\section*{\centerline{IV.NUMERICAL EXPERIMENTS}}

In this section, we present the numerical results of our proposed
algorithm and compare its performance with the state-of-the art algorithms.
As CAN algorithm and ADMM method developed in \cite{19_ADMM} solves
an approximate problem, we will not include them for numerical comparision.
So, we compared our results with MISL and ISL-NEW algorithm. All the
simulations were performed in MATLAB on a PC with two core 2.40GHz
processor. Experiments has been conducted to design a sequence of
lengths $N=50,100,200,300,400,500$ using different initialization
sequences like Random, Golomb \cite{15_polyphaseseq_Zhang} and Frank
\cite{13_Polyphasecode_frank} sequences. In case of random initialization,
$30$ monte carlo runs has been conducted for every length and for
every run, initialization sequence $\bigl\{ y_{i}^{0}\bigr\}_{i=1}^{N}$
is choosen as $\bigl\{ e^{j2\pi\theta_{i}}\bigr\}_{i=1}^{N}$, where
$\bigl\{\theta_{i}\bigr\}$ are drawn randomly from the uniform distribution
$\left[0,1\right]$. The convergence criterion which we used to stop
all the algorithms in the comparision is

\begin{equation}
\Biggr|\frac{(\text{ISL}(t+1)-\text{ISL}(t))}{\text{max}(1,\text{ISL}(t))}\Biggr|\leq10^{-5},\label{eq:40}
\end{equation}

where $\text{ISL}(t)$ is the ISL metric value at $t^{th}$ iteration.

In each experiment, execution time of the proposed algorithm and property
of a designed sequence such as auto-correlation side-lobe levels,
cost function value were observed and compared with the MISL and ISL-NEW
algorithms. Since the algorithms under comparision, MISL and ISL-NEW
and also our algorithm are based on MM, all of them can be accelerated
using standard acceleration schemes \cite{26_ACC_varadhan}, \cite{27_ACC_Raydan},
\cite{28_ACC_barzilai}, but for the sake of comparison, we didn't
implement acceleration scheme for any of the methods.

\begin{figure}[tp]
\subfloat[$N=100$]{\includegraphics[width=8.2cm,height=5.9cm]{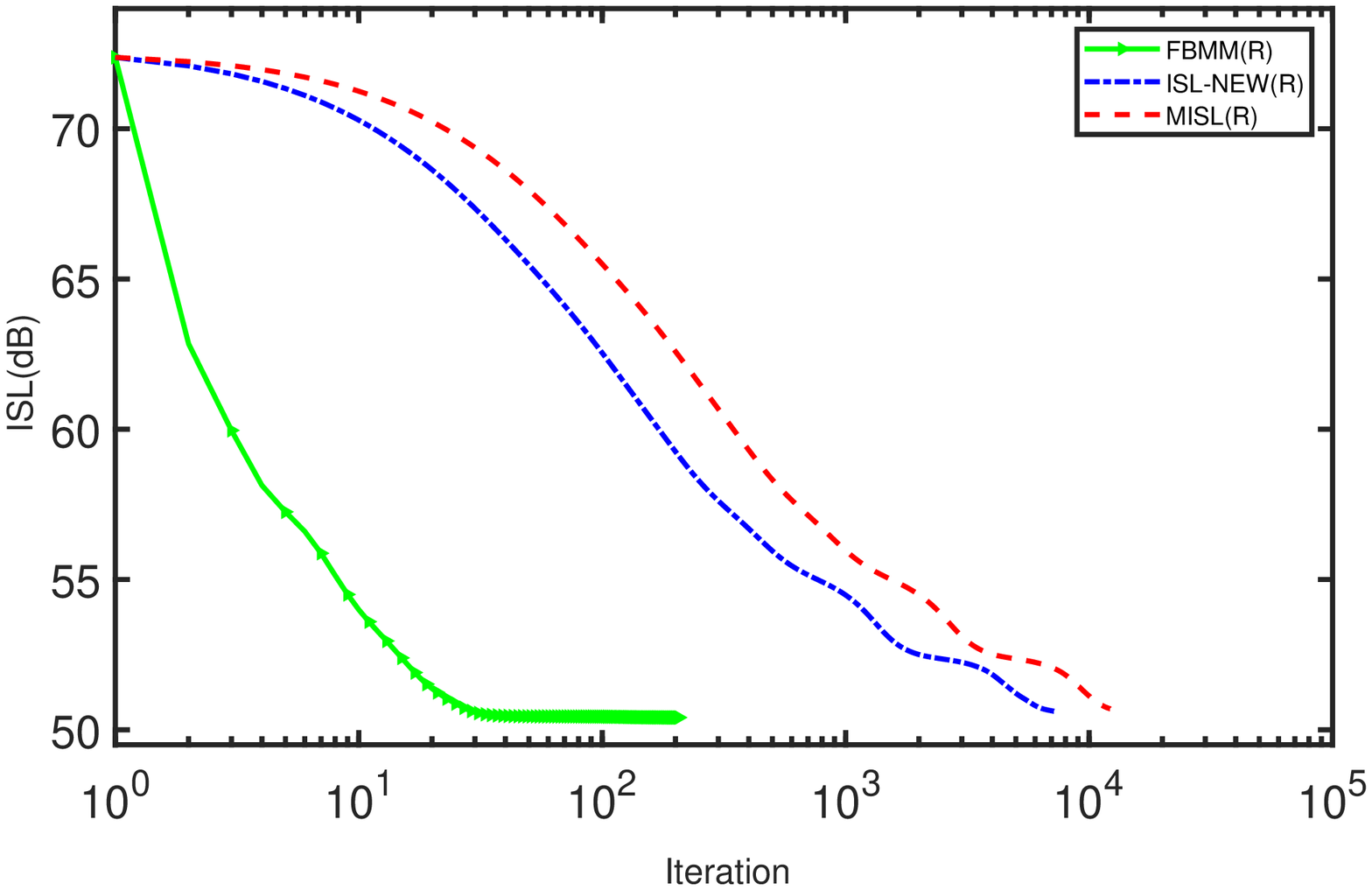}}\subfloat[$N=500$]{\includegraphics[width=8.2cm,height=5.9cm]{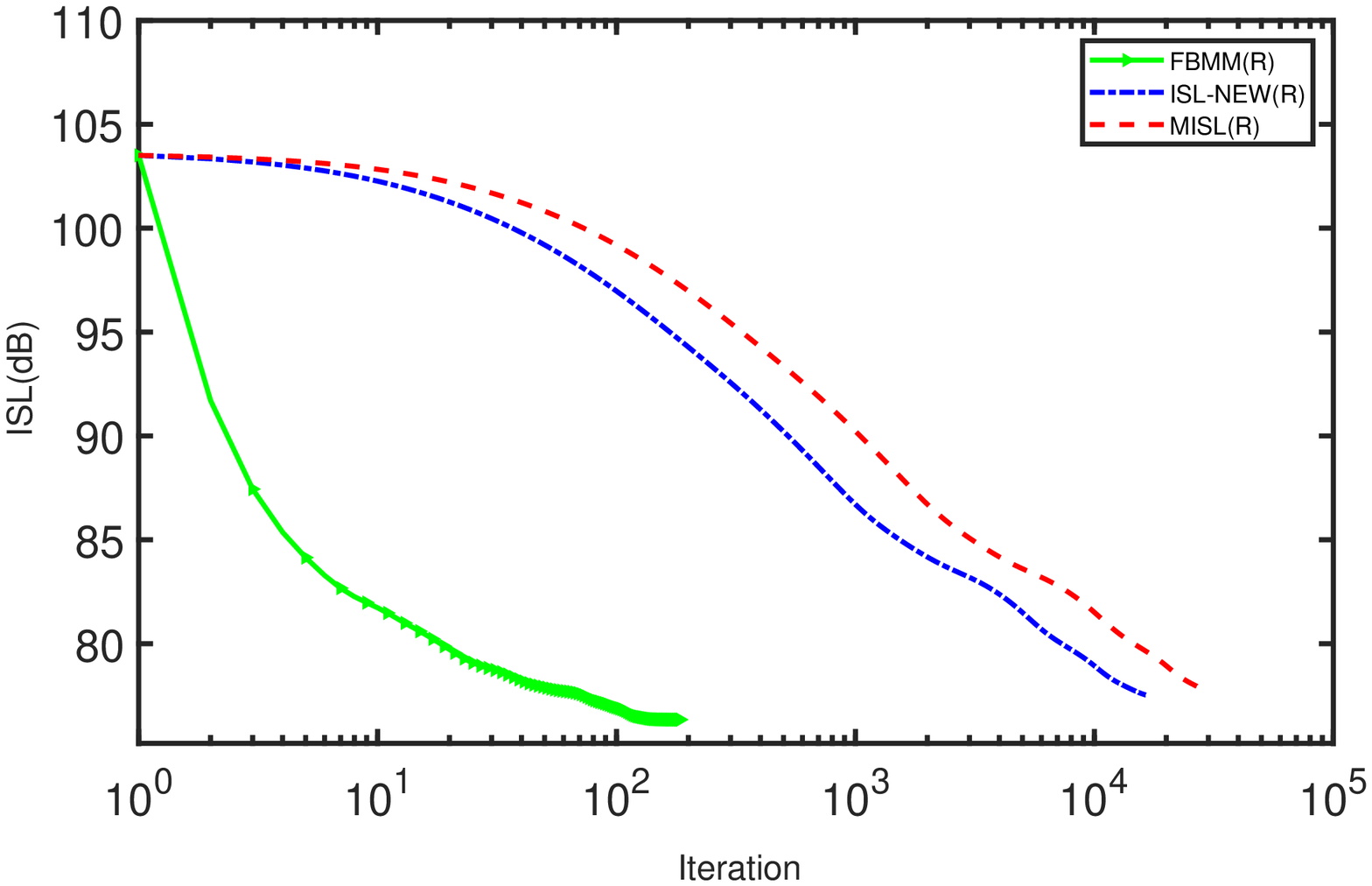}}

\subfloat[$N=100$]{\includegraphics[width=8.2cm,height=5.9cm]{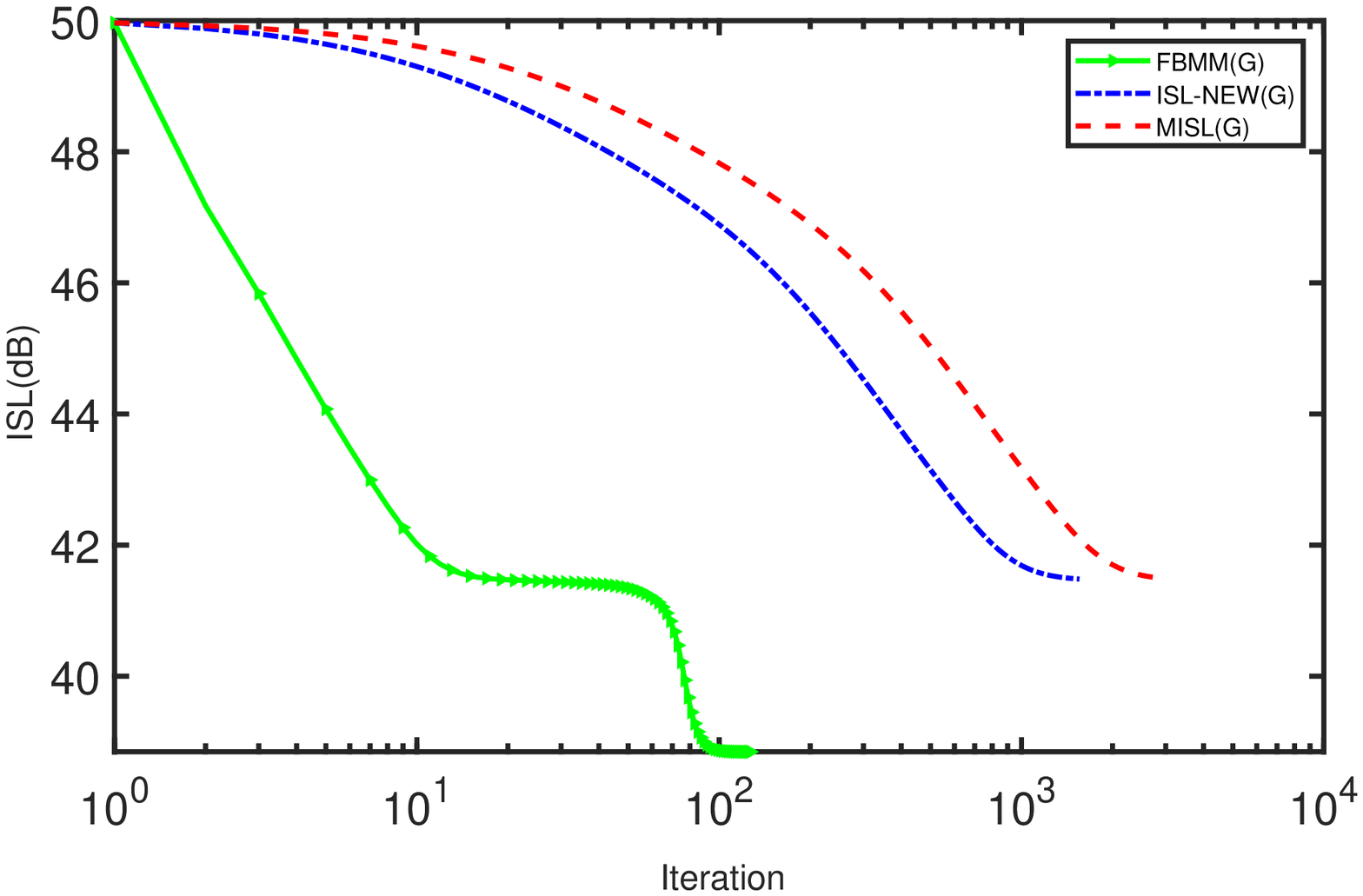}}\subfloat[$N=500$]{\includegraphics[width=8.2cm,height=5.9cm]{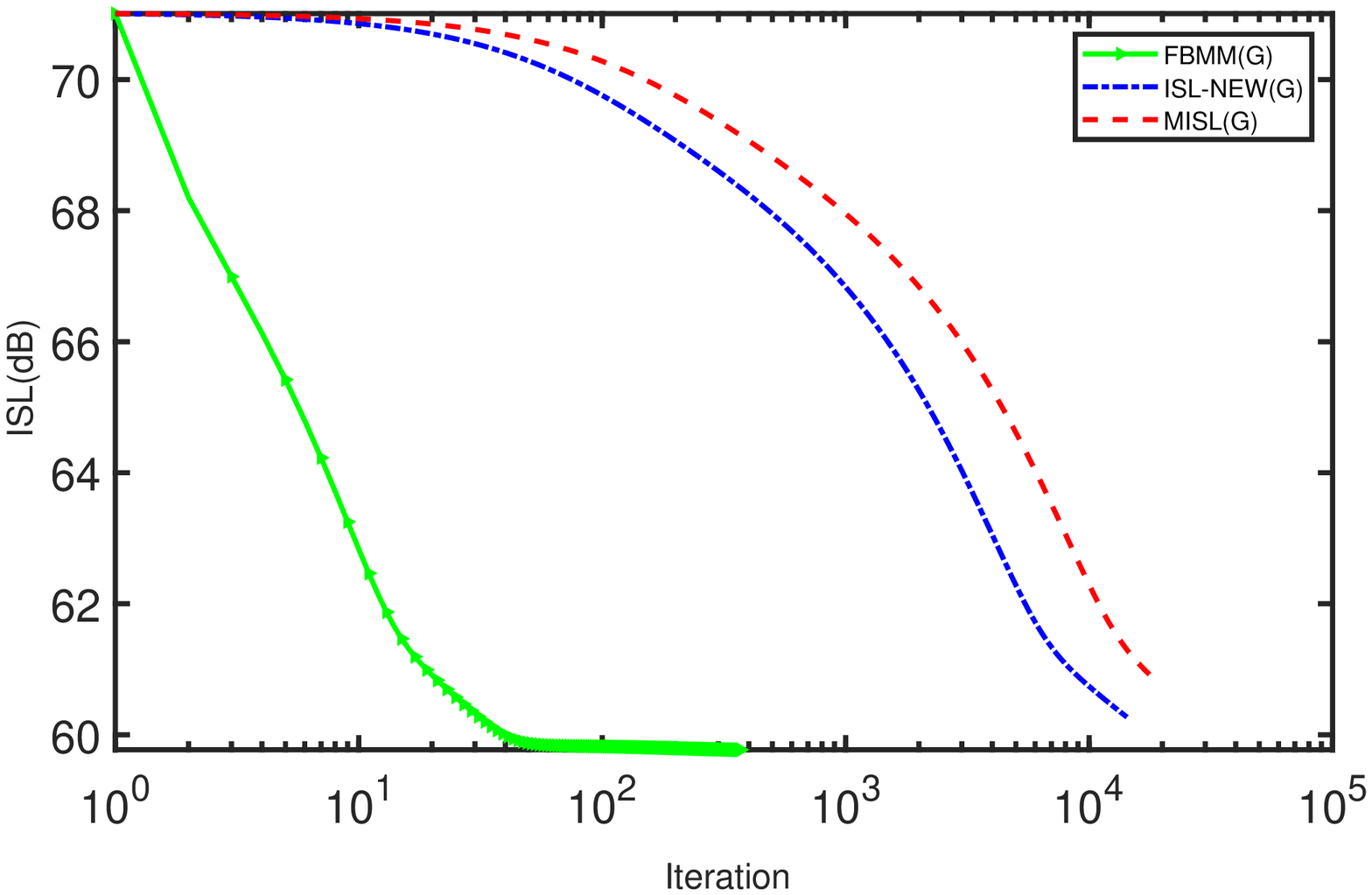}}

\subfloat[$N=289$]{\includegraphics[width=8.2cm,height=5.9cm]{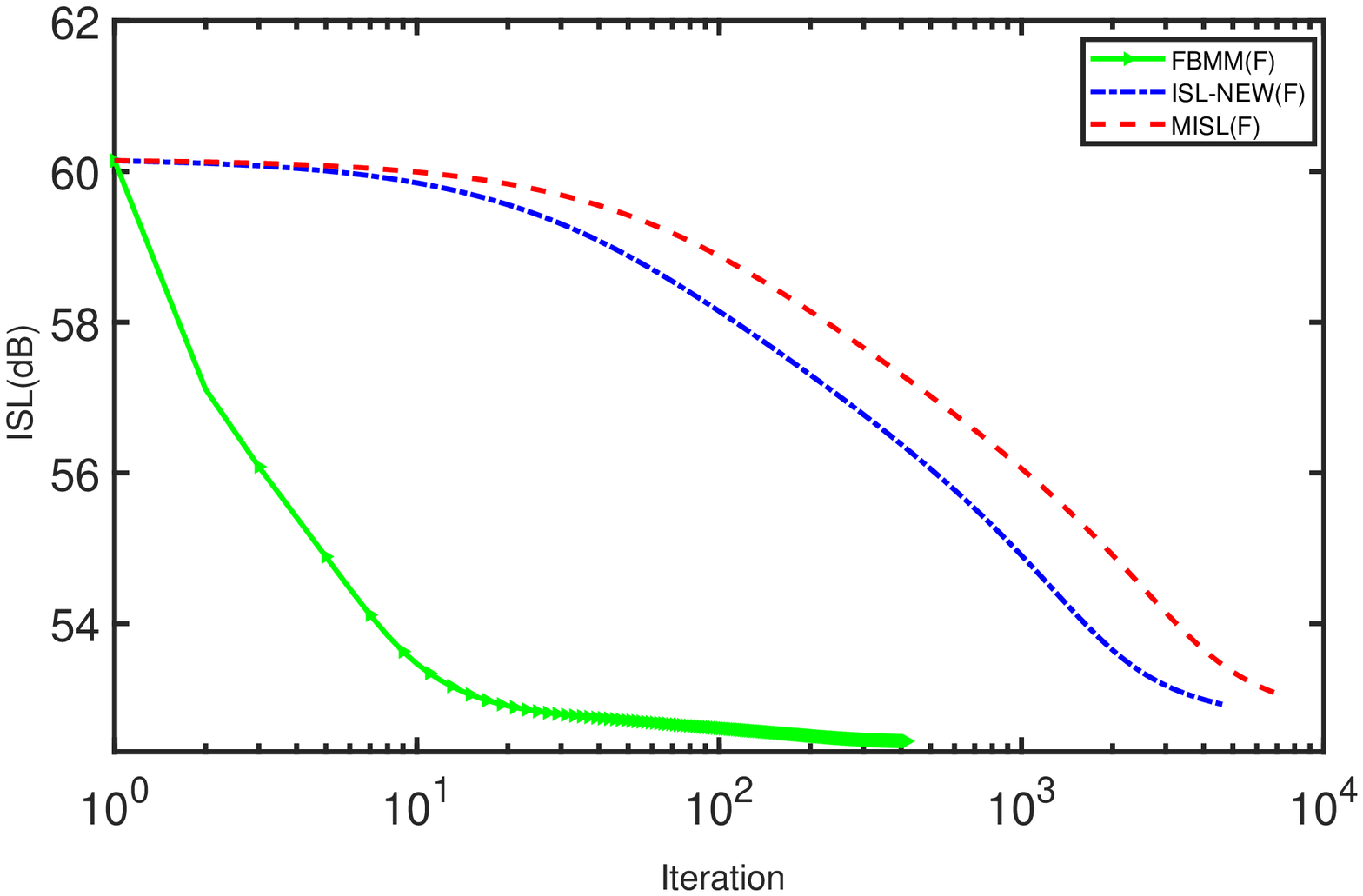}}\subfloat[$N=484$]{\includegraphics[width=8.2cm,height=5.9cm]{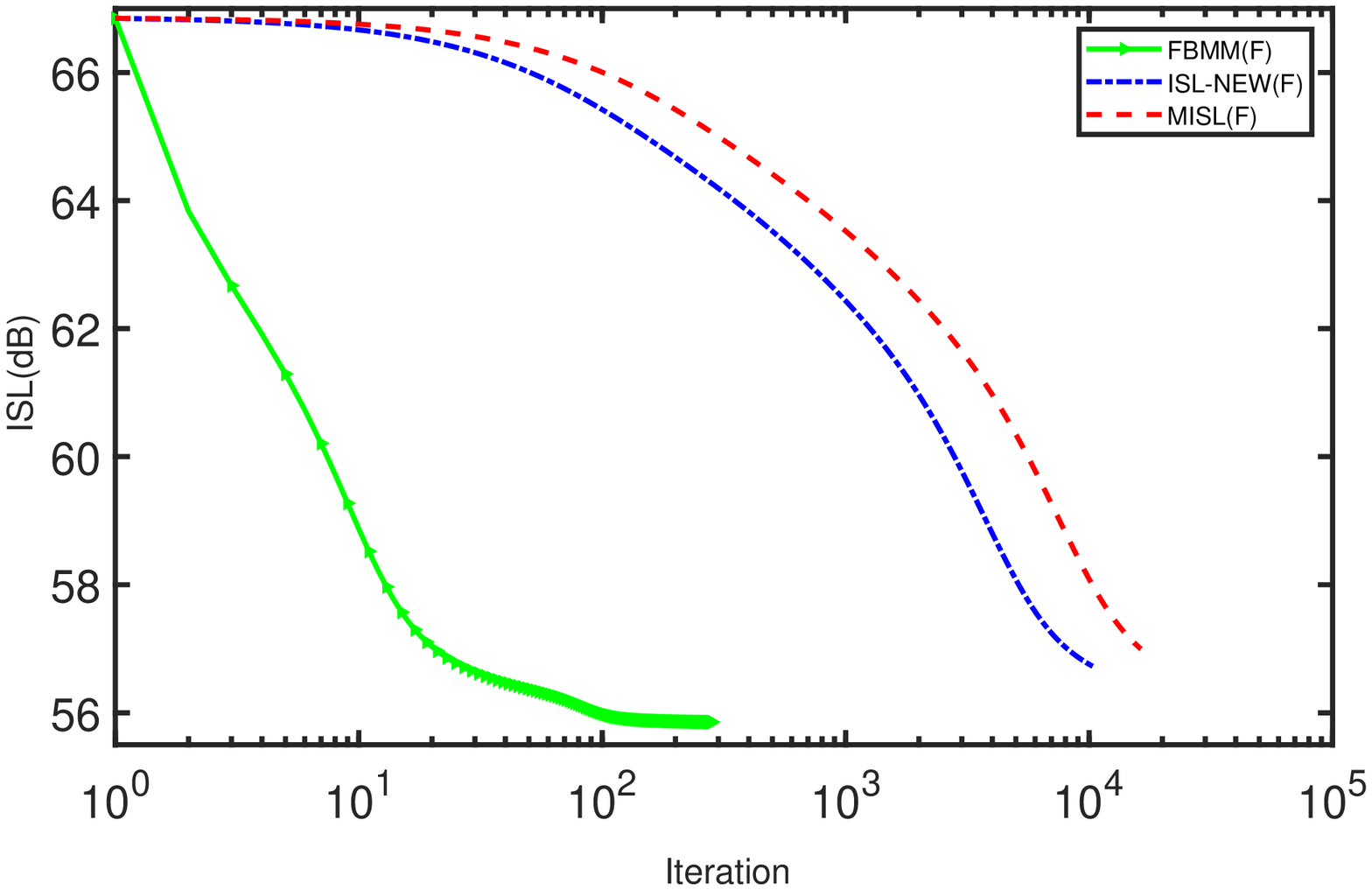}}

\caption{ISL vs Iteration for a sequence length $N=100,289,484,500$. (a) and
(b) are for initialization via Random sequence. (c) and (d) are for
initialization via Golomb sequence. (e) and (f) are for initialization
via Frank sequence.}
\end{figure}

Figure. 1 consists the plots of ISL value vs iteration number for
different lengths using three different initialization sequences.
Here FBMM(R), FBMM(G), FBMM(F) indicates FBMM algorithm initialized
with Random, Golomb and Frank sequences, respectively. We initialize
all the algorithms at the same initial point and observed that, almost
they also ended up at the same minimum but with different speed of
convergence. From the plots, it can be observed that for the different
lengths $N$, MISL and ISL-NEW are taking much larger number of iterations
and FBMM is taking lesser number of iterations to converge to the
same objective minimum value.

\begin{figure}[tp]
\subfloat[$N=100$]{\includegraphics[scale=0.55]{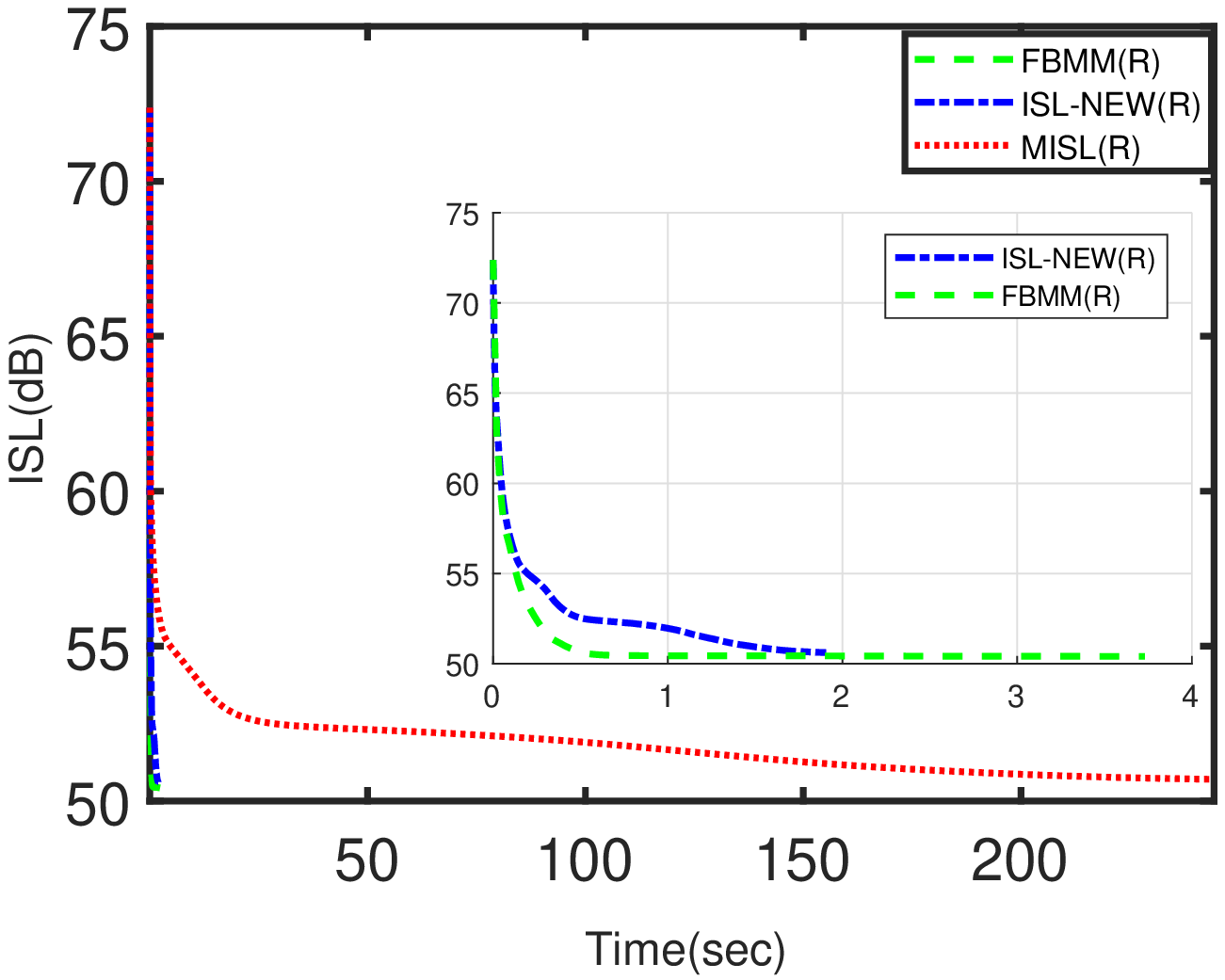}}\subfloat[$N=500$]{\includegraphics[scale=0.55]{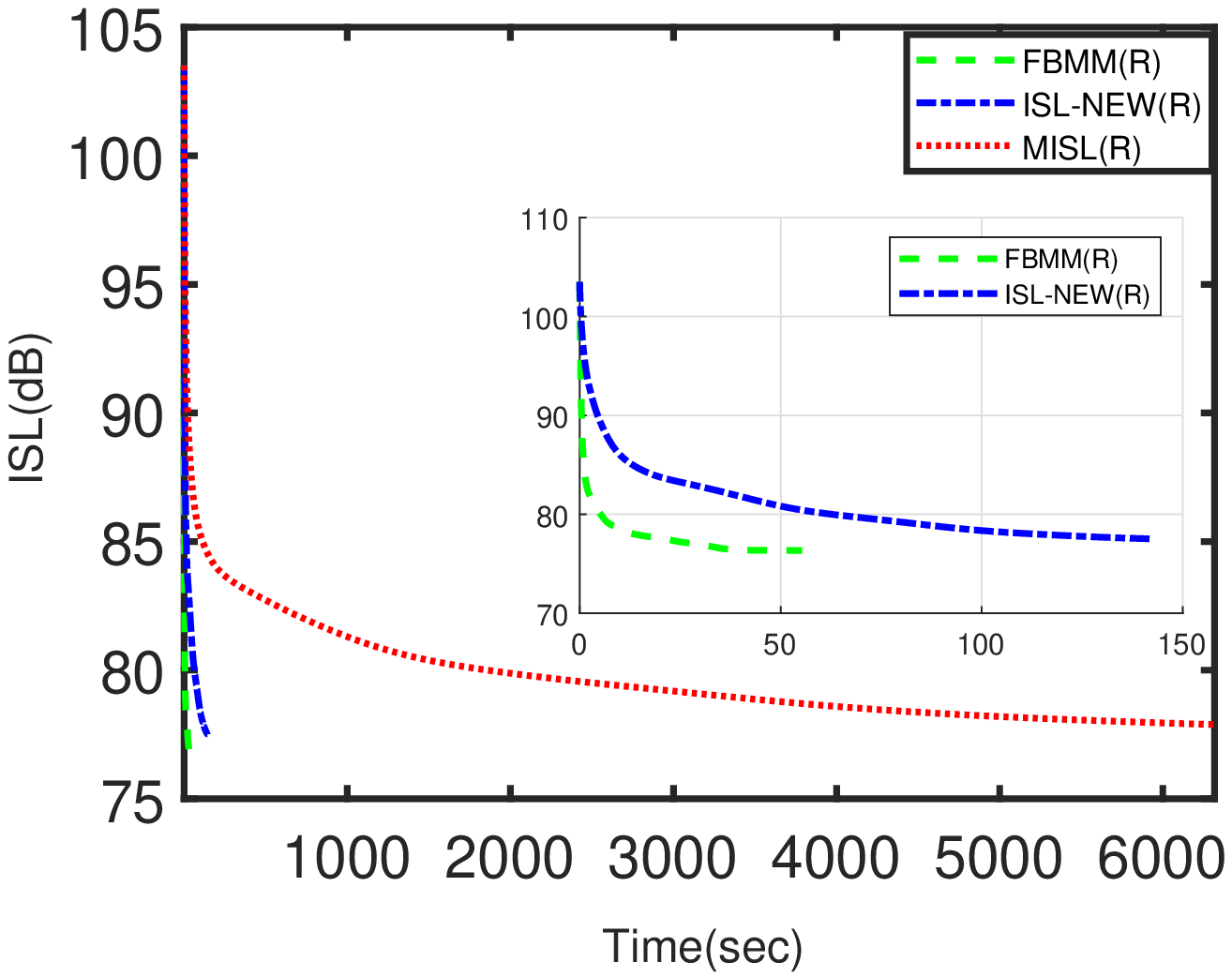}}

\subfloat[$N=100$]{\includegraphics[scale=0.55]{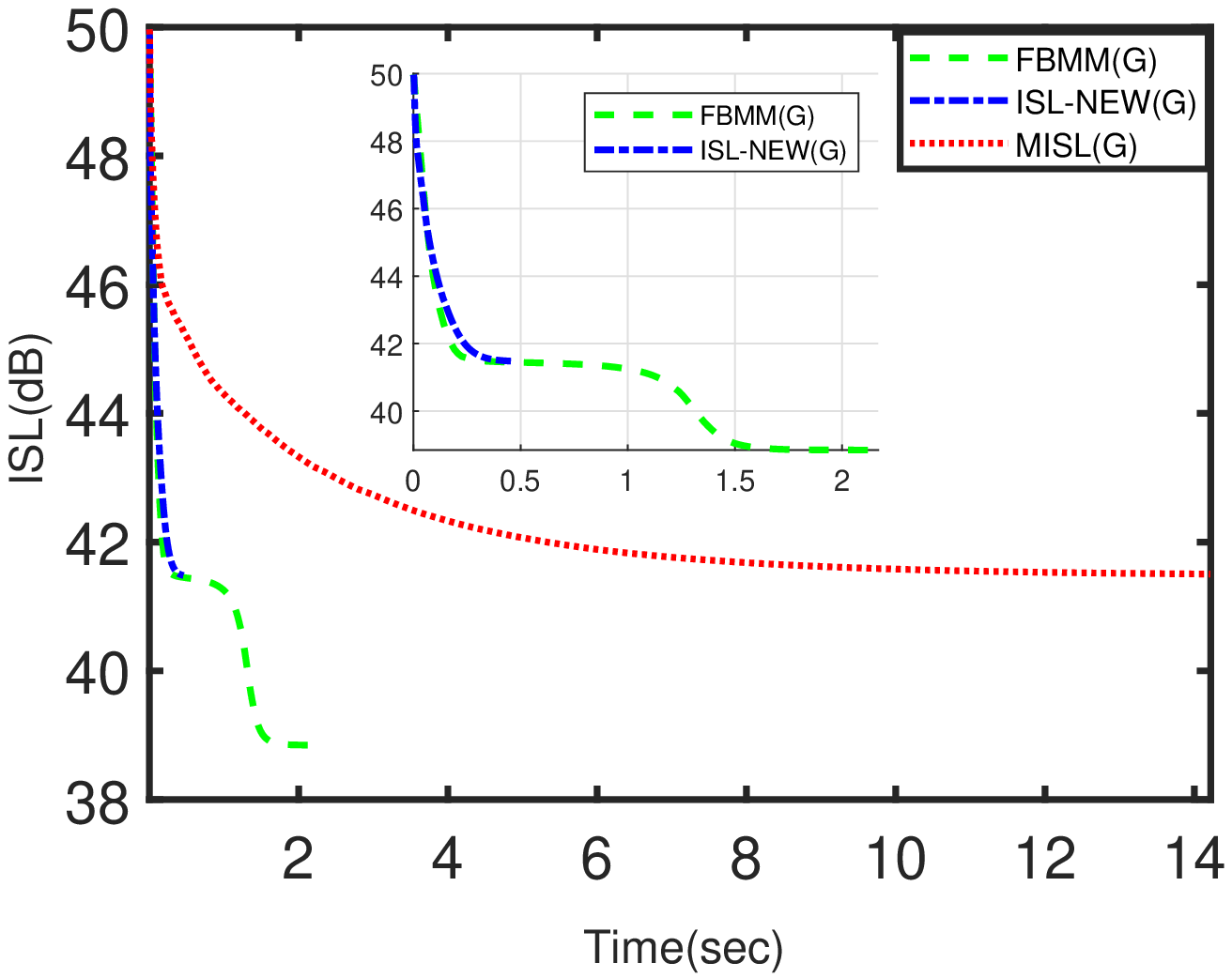}}\subfloat[$N=500$]{\includegraphics[scale=0.55]{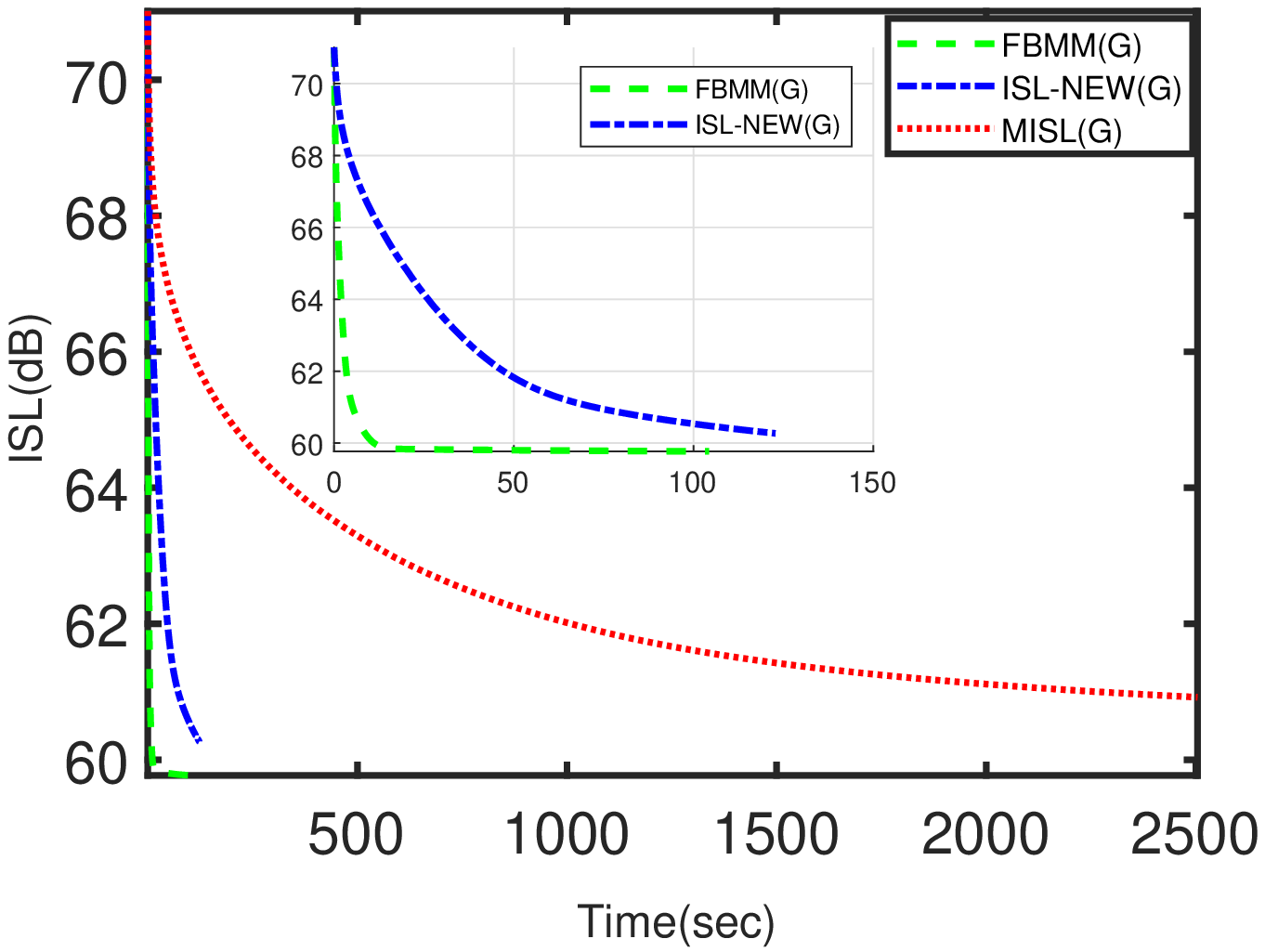}}

\subfloat[$N=289$]{\includegraphics[scale=0.55]{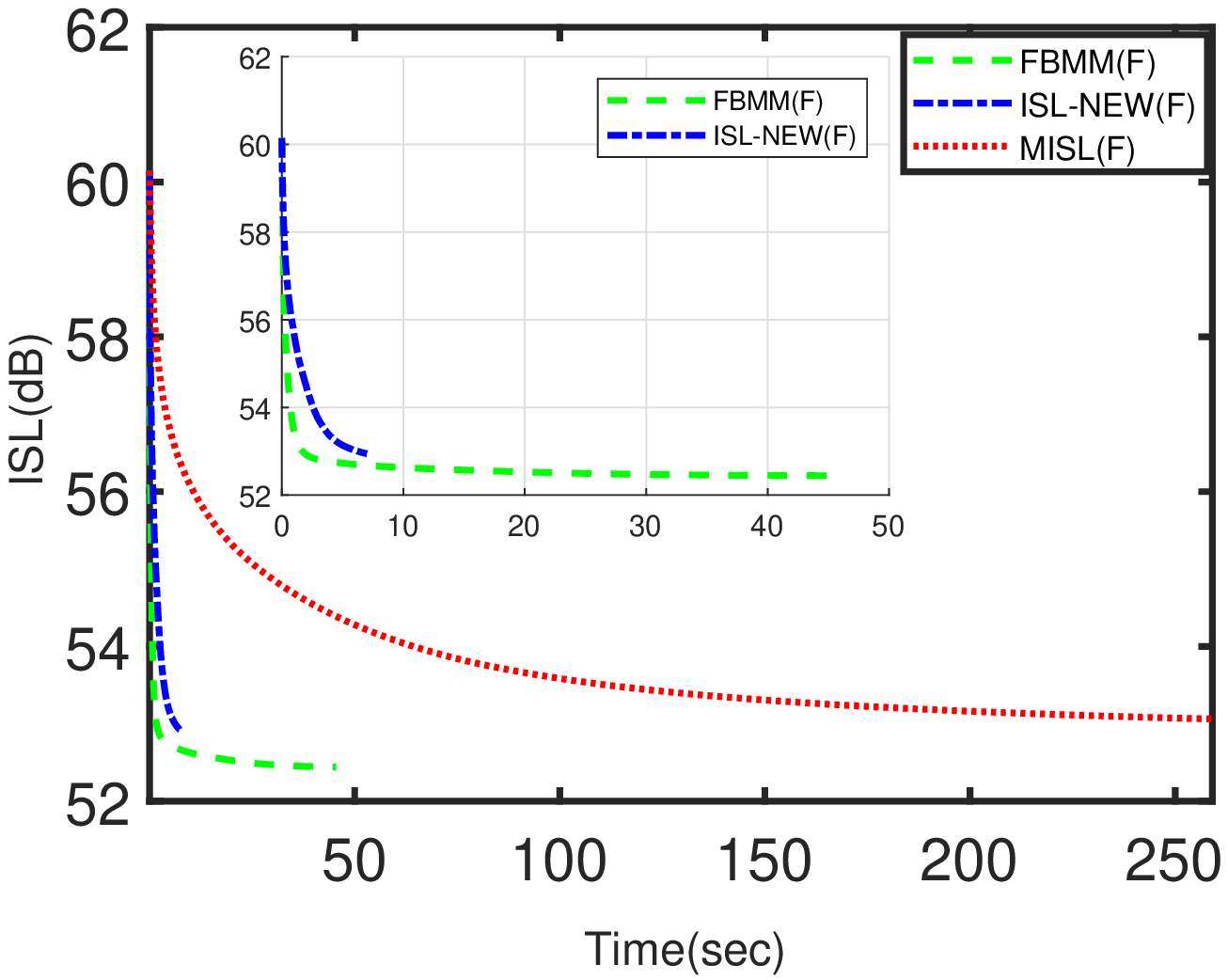}}\subfloat[$N=484$]{\includegraphics[scale=0.55]{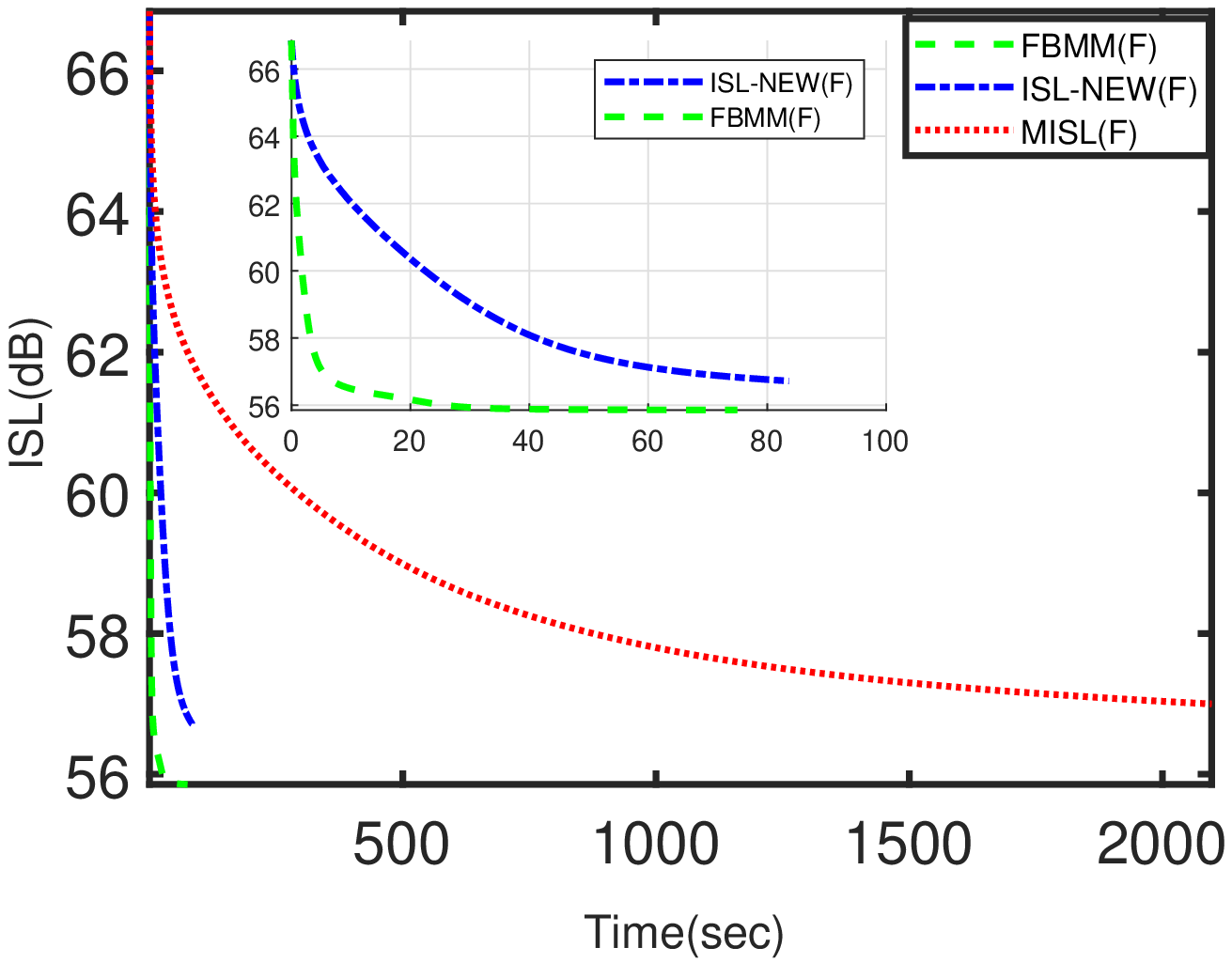}}

\caption{ISL vs Time for a sequence length $N=100,289,484,500$. (a) and (b)
are for initialization via Random sequence. (c) and (d) are for initialization
via Golomb sequence. (e) and (f) are for initialization via Frank
sequence.}
\end{figure}

Figure. 2 shows the ISL value at every iteration vs time for different
lengths using different initialization sequences. Each plot also has
zoomed version to show the subtle difference in the speed of convergence
of FBMM and ISL-NEW. From the plots, it can be observed that, irrespective
to the initialization sequence and length $N$, FBMM is always taking
less time to converge to the same objective minimum value when compared
to MISL and ISL-NEW algorithms.

\begin{figure}[tp]
\subfloat[$N=100$]{\includegraphics[scale=0.55]{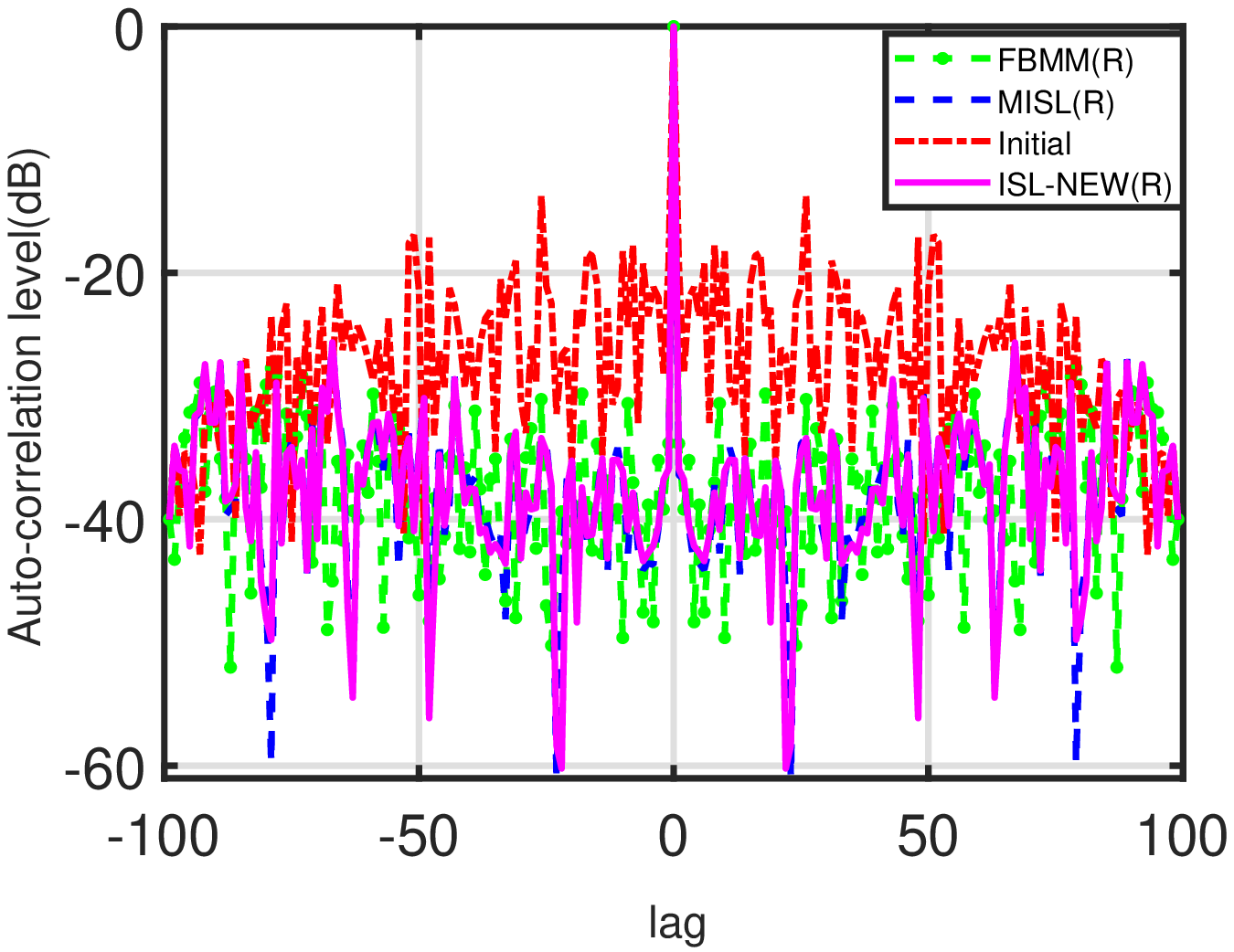}}\subfloat[$N=500$]{\includegraphics[scale=0.55]{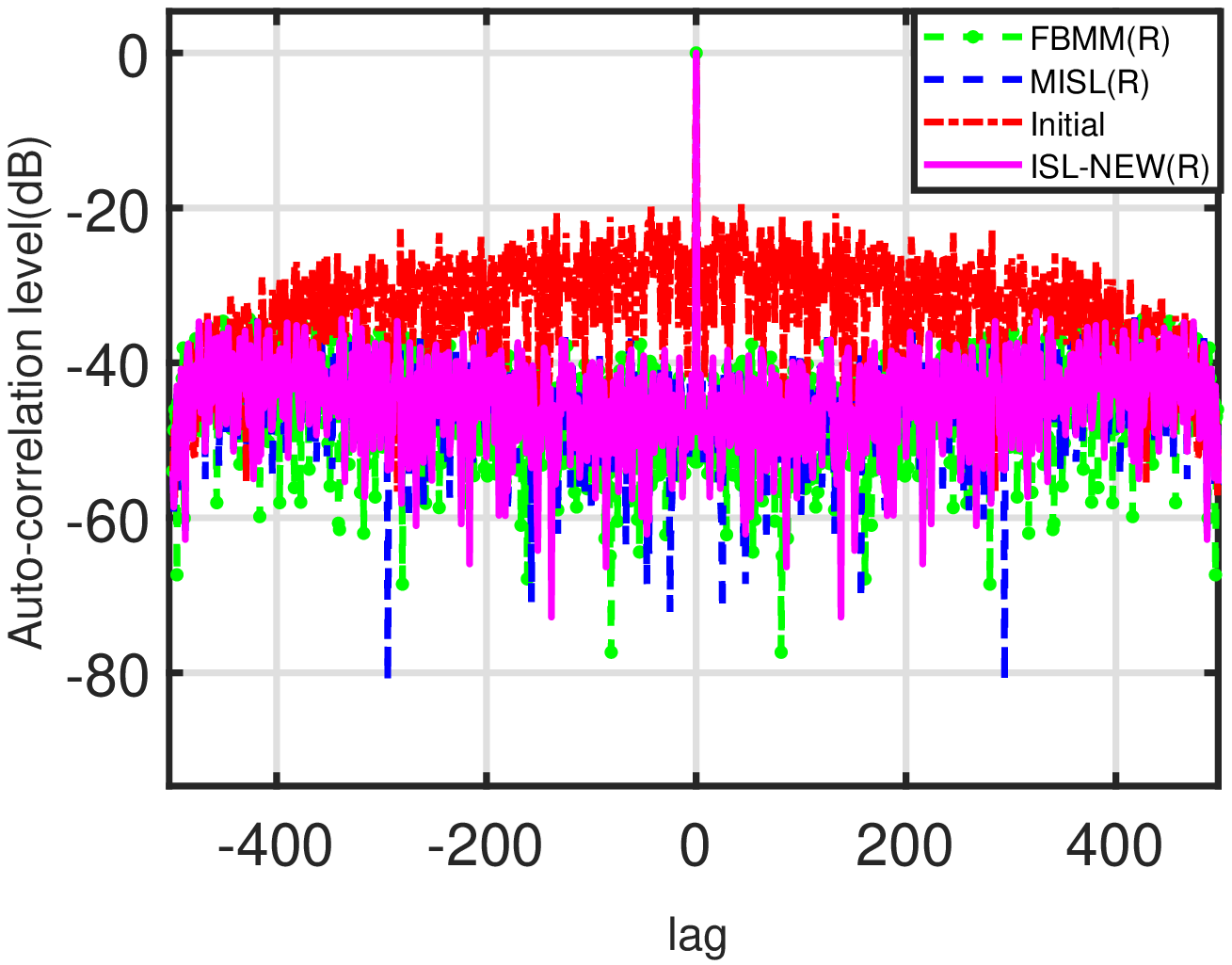}}

\subfloat[$N=100$]{\includegraphics[scale=0.55]{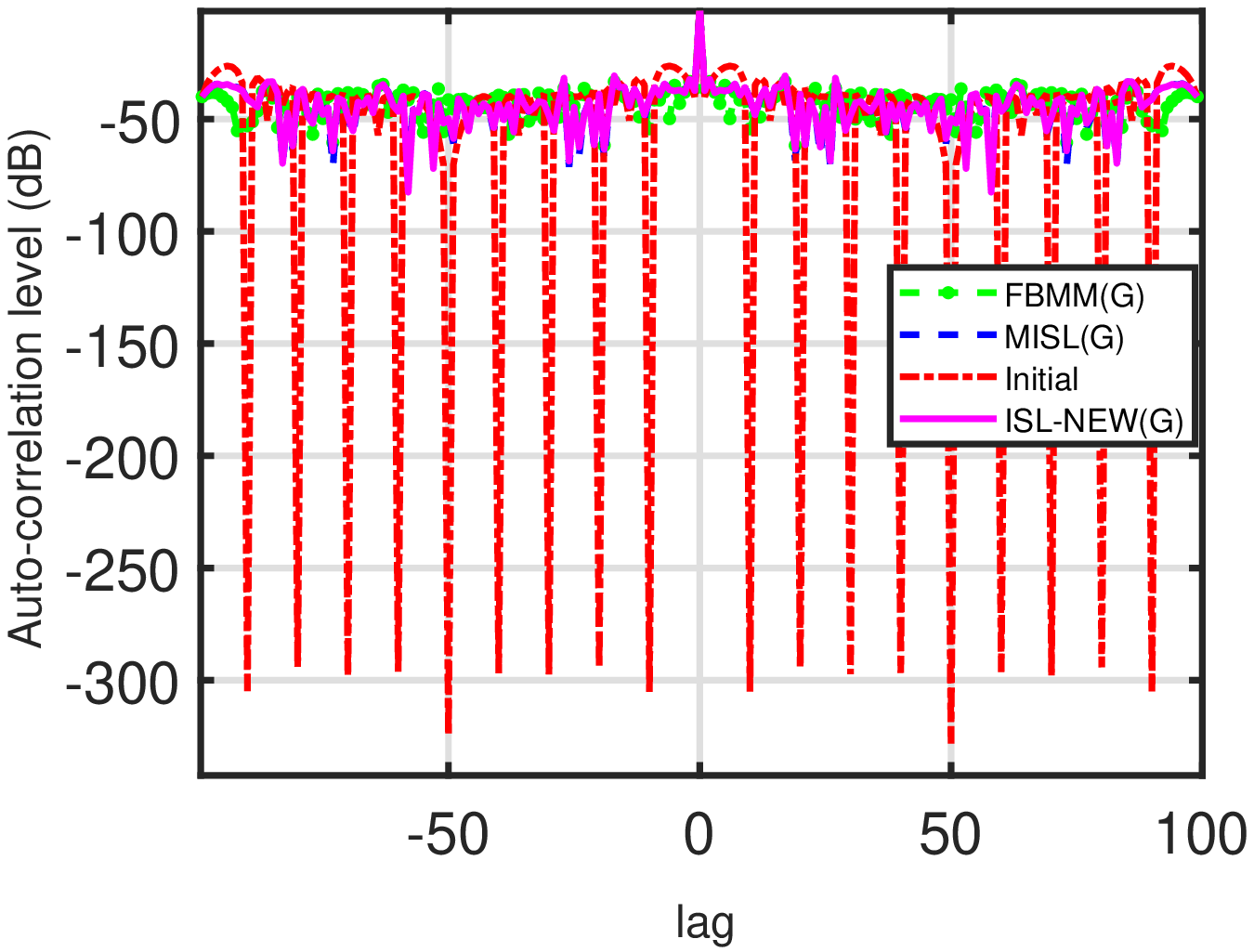}}\subfloat[$N=500$]{\includegraphics[scale=0.55]{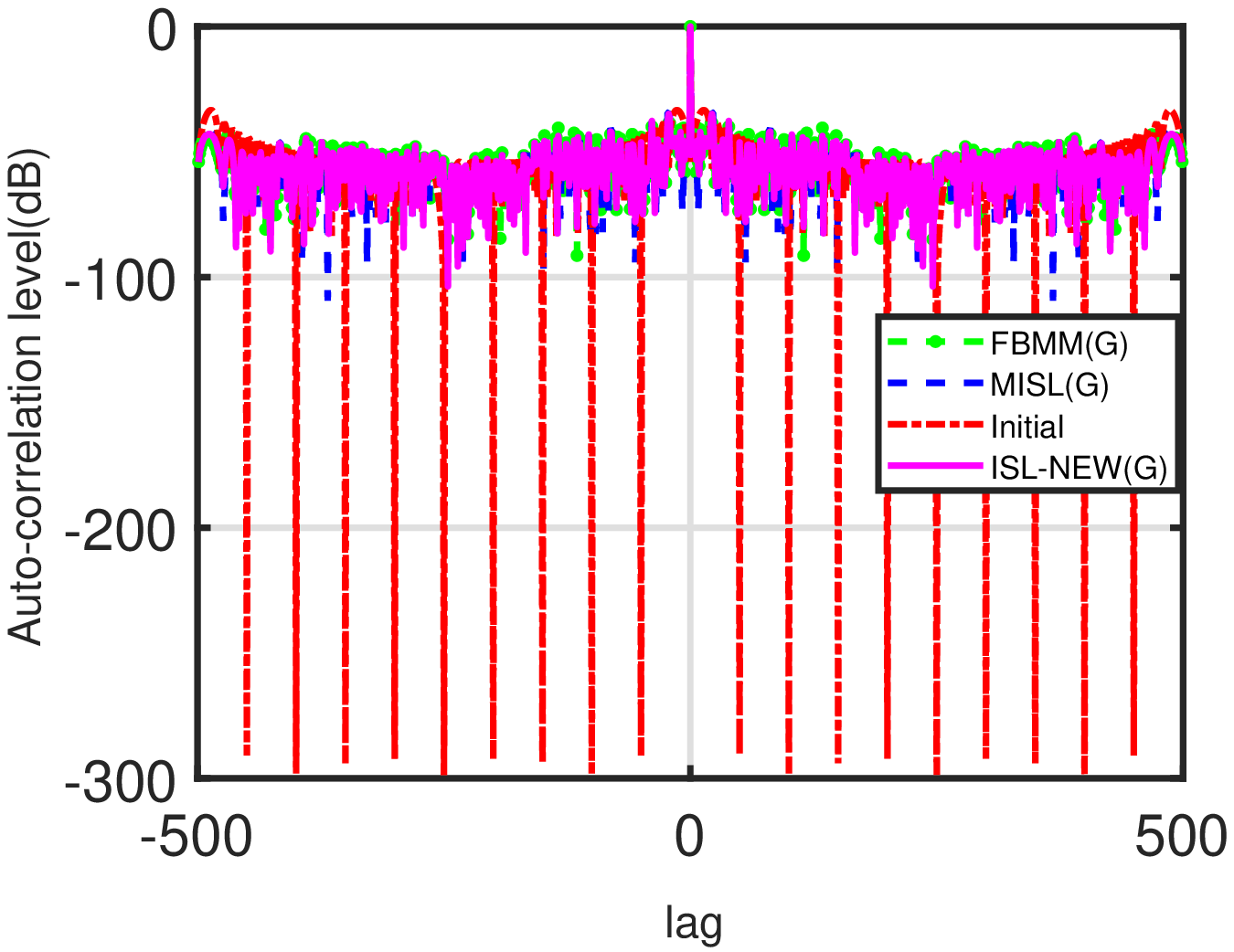}}

\subfloat[$N=289$]{\includegraphics[scale=0.55]{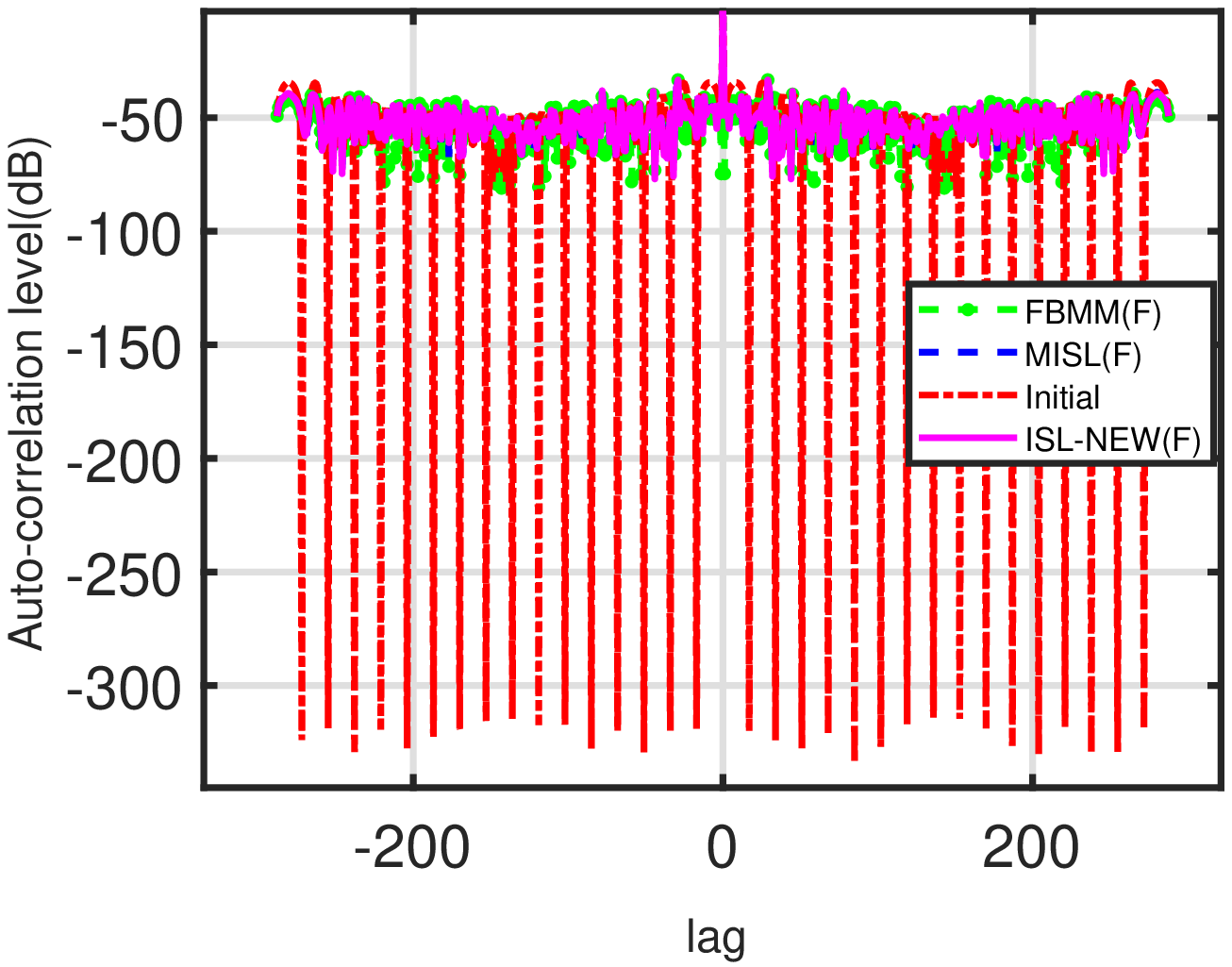}}\subfloat[$N=484$]{\includegraphics[scale=0.55]{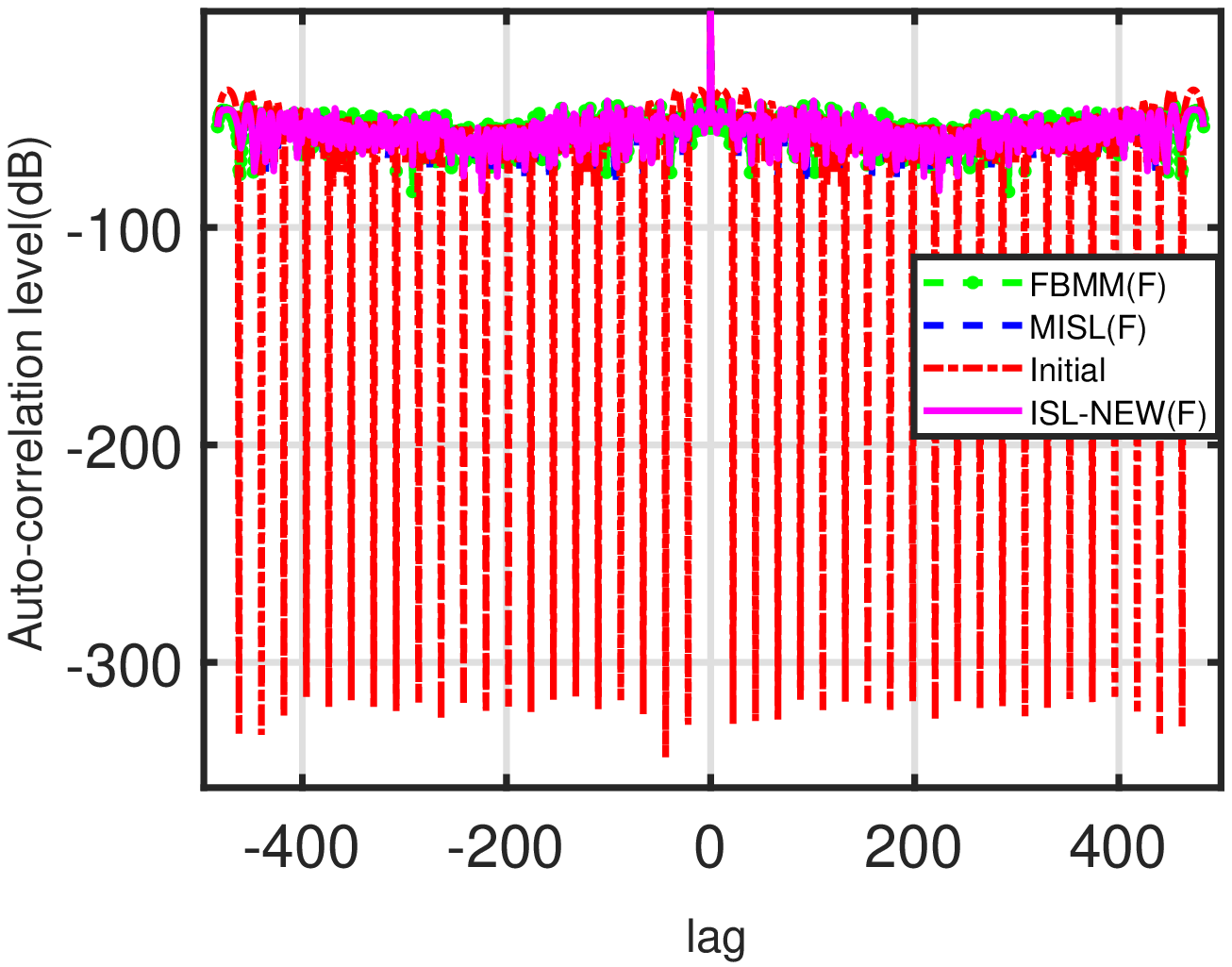}}

\caption{Autocorrelation value vs lag for a sequence length $N=100,289,484,500$.
(a) and (b) are for initialization via Random sequence. (c) and (d)
are for initialization via Golomb sequence. (e) and (f) are for initialization
via Frank sequence.}
\end{figure}

Figure. 3 shows the auto-correlation plots of the generated sequence
via FBMM, ISL-NEW and MISL algorithms using different initialization
sequences. From plots, we observed that in the case of Frank and Golomb
sequence initializations, most of the side-lobe levels in the initial
sequence itself are too low but PSL is high. It can be seen that all
the three algorithms improves performance from the initialized sequence
interms of PSL.

\begin{figure}[tp]
\begin{centering}
\includegraphics[scale=0.55]{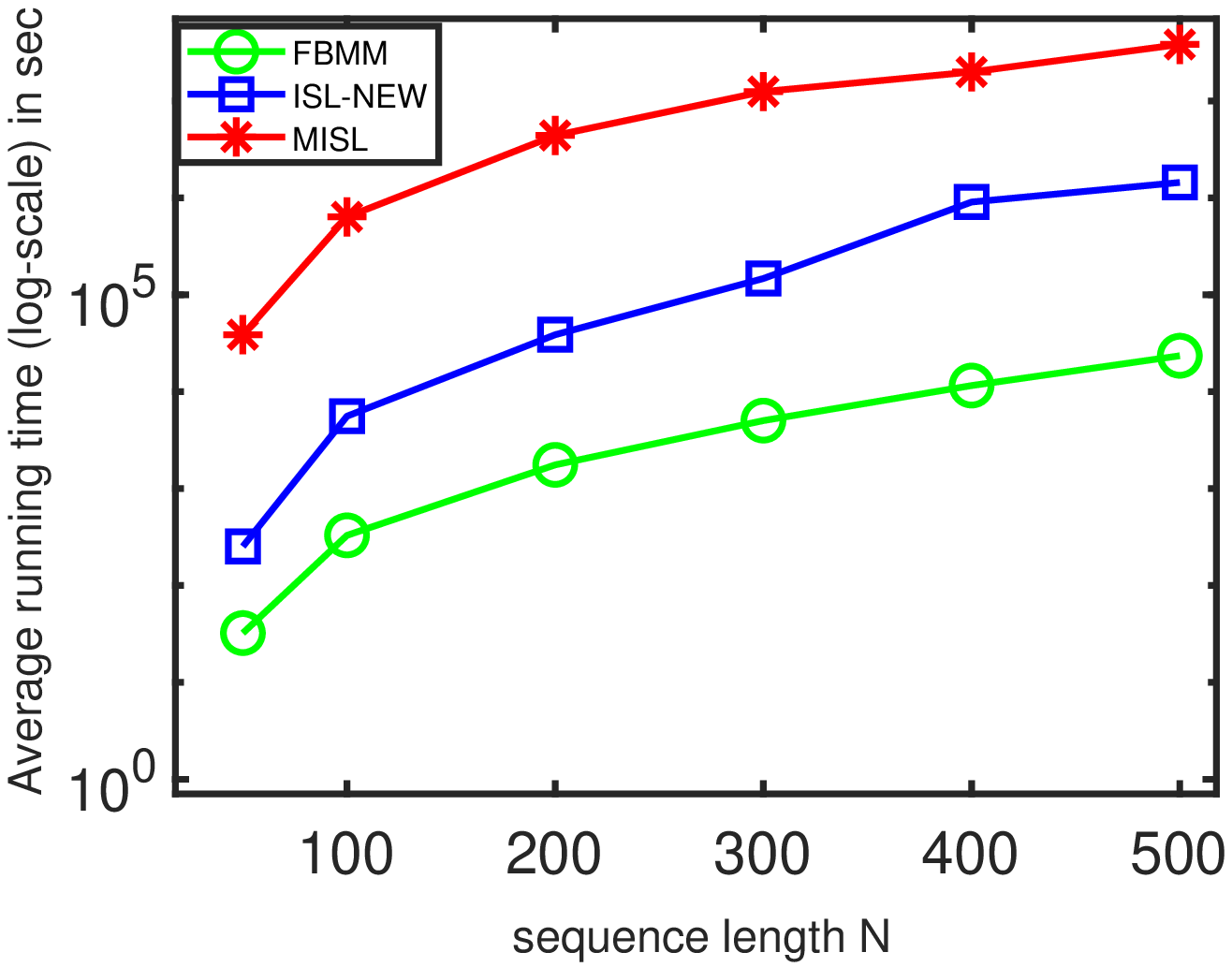}
\par\end{centering}
\centering{}\caption{Avg-Time vs sequence length using Random initialization sequence.}
\end{figure}

Figure. 4 has a comparision of three algorithms interms of average
running time for different lengths using random initialization sequence.
For better comparision, all the three algorithms are initialized with
a same sequence and stopped using the same convergence criterion.
From the figure, it can be observed that, irrespective of length $N$,
FBMM is taking lesser time to converge when compared to rest of the
two algorithms: MISL and ISL-NEW.

\section*{\centerline{VI.Conclusion}}

To design a sequence of any length $(N)$, we have proposed an algorithm
by minimizing the ISL metric. We also shown a computationally efficient
way of implementing our proposed algorithm and named as FBMM. Proposed
algorithm is derived based on a Block MM technique and implemented
using FFT, IFFT operations, hence computationally efficient for large
lengths. Numerical experiments shows that, proposed algorithm is performing
well when compared to state-of-the art algorithms in terms of convergence
rate, computational complexity and average running time.

\bibliographystyle{IEEEtran}
\nocite{*}
\bibliography{FBMM_REFERENCE_FINAL}

\end{document}